\documentclass[twocolumn,showpacs,amsmath,amssymb,prd,showkeys,10pt,nofootinbib,superscriptaddress]{revtex4}

\usepackage{epsfig}
\usepackage{latexsym}
\usepackage{graphicx}
\usepackage{amsfonts}
\usepackage{amsmath}
\usepackage{natbib}
\usepackage{xcolor}
\usepackage{wasysym}
\usepackage{bm}
\usepackage{array}

\newcommand{\etal}{{et~al.}}
\newcommand{\sz}{{\sc sz}}
\newcommand{\zb}{{\sc zb}}
\newcommand{\kn}{{\sc kn}}

\def\apj{Astophys. J.}%
\def\aap{Astron. \& Astrophys.}%
\def\prd{Phys.~Rev.~D}%
\def\prl{Phys.~Rev.~Lett.}%
\def\nat{Nature}%
\def\l#1{\left#1}
\def\r#1{\right#1}
\def\ds{\displaystyle}
\def\eref#1{(\ref{#1})}

\def\simlt{\lower.5ex\hbox{$\; \buildrel < \over \sim \;$}}
\def\simgt{\lower.5ex\hbox{$\; \buildrel > \over \sim \;$}}

\begin{document}
\title{Efficiency of pseudo-spectrum methods for estimation of Cosmic Microwave Background $B$-mode power spectrum}

\author{ A. Fert\'e}
\email{agnes.ferte@ias.u-psud.fr}
\affiliation{%
 Universit\'e Paris-Sud 11, Institut d'Astrophysique Spatiale, UMR8617, Orsay, France, F-91405}
\affiliation{%
CNRS, Orsay, France, F-91405}

\author{J. Grain}
 \email{julien.grain@ias.u-psud.fr}
\affiliation{%
CNRS, Orsay, France, F-91405}
 \affiliation{%
 Universit\'e Paris-Sud 11, Institut d'Astrophysique Spatiale, UMR8617, Orsay, France, F-91405}

\author{M. Tristram}%
 \email{tristram@lal.in2p3.fr}
\affiliation{%
CNRS, Orsay, France, F-91405}
\affiliation{Universit\'e Paris-Sud 11, Laboratoire de l'Acc\'el\'erateur Lin\'eaire, B\^atiment 200, 91898 Orsay Cedex, France}

 \author{R. Stompor}%
 \email{radek@apc.univ-paris-diderot.fr}
\affiliation{%
AstroParticule et Cosmologie, Universit\'e Paris Diderot, CNRS/IN2P3, CEA/Irfu, Obs. de Paris, Sorbonne Paris Cit\'e, France 
}

\begin{abstract}
Estimation of the $B$-mode angular power spectrum of polarized anisotropies of the cosmic microwave background (CMB) is a key step towards a full exploitation of the scientific potential of this probe.
In the context of pseudo-spectrum methods the major challenge  is related to a contamination of the $B$-mode spectrum estimate with residual power of much larger $E$-mode. This so-called $E$-to-$B$ 
leakage is unavoidably present whenever an incomplete sky map is only available, as is the case for any realistic observation.
The leakage has to be then minimized or removed  and ideally in such a way that neither a bias nor extra  variance is
introduced. In this paper, we compare from these two perspectives three different methods proposed recently in this context \cite{smith_2006, zhao_baskaran_2010, kim_naselsky_2010},
which we first introduce within a common algebraic framework of the so-called $\chi$-fields and then study their performance on two different experimental configurations -- one corresponding to a small-scale experiment 
covering 1\% of the sky motivated by current ground-based or balloon-borne experiments and another -- to a nearly full-sky experiment, {\it e.g.}, a possible CMB $B$-mode satellite mission.
We find that though all these methods allow to reduce significantly the level of the $E$-to-$B$ leakage,  it is the method of \cite{smith_2006}, which at the same time ensures the smallest error bars in all experimental configurations studied here, owing to the fact that it permits straightforwardly
for an optimization of the sky apodization of the polarization maps used for the estimation. For a satellite-like experiment, this method enables a detection of $B$-mode power spectrum at large angular scales but only after appropriate binning.
The method of \cite{zhao_baskaran_2010} is a close runner-up in the case of a nearly full sky coverage. 
\end{abstract}

\pacs{98.80.-k; 98.70.Vc; 07.05.Kf}
\keywords{Cosmology: cosmic background radiation--Cosmology: observation}

\maketitle


\section{Introduction}
\label{sect:intro}

Polarized anisotropies of the cosmic microwave background (CMB) radiation come in two flavors: gradient-like, $E$, component, and curl-like, $B$, \cite{zaldarriaga_seljak_1997,kamionkowski_etal_1997}.
Ten years ago, the first detection of the $E$-mode anisotropies  was announced by the \textsc{dasi} team \cite{dasi}. 
Since then many subsequent experiments {\it e.g.} {\sc wmap} \cite{wmap_pol}, {\sc quad} \cite{quad_pol}, or {\sc bicep} \cite{bicep_pol} have detected the $E$-mode anisotropies  with high significance deepening and confirming our understanding of the Universe's evolution and structure formation. \textsc{planck}~\cite{planck}  is widely expected to provide shortly most comprehensive and precise constraints on the $E$-mode polarization properties in a range of angular scales extending from the largest down to few arc minutes.

In contrast, no $B$-mode anisotropy has been detected yet only some upper limits are currently available, {\it e.g.},  \cite{wmap_pol,quad_pol,bicep_pol}. This is expected given minute amplitudes predicted for this signal. 
At the same time the scientific potential of the $B$-mode probe has been generally recognized as extremely promising.
For instance, on the linear level the $B$-modes can be sourced by the primordial gravitational waves \cite{seljak_zaldarriaga_1997,spergel_zaldarriaga_1997} and not by the scalar fluctuations, thought to be largely responsible for the observed total intensity and $E$-mode 
anisotropies. Consequently,  a detection of the $B$-mode anisotropy at large angular scales ($\ell \apprle 100$) in excess of what is expected from the gravitational lensing signal, see below, could be seen as a direct validation of inflationary theories, as the latter are considered to be the most likely source of the gravity waves, and could allow for discrimination between different inflationary models. It could also set useful constraints on the reionization period \cite{zaldarriaga_1997}.
At smaller angular scales, $B$-modes are expected to be mainly due to gravitational lensing of CMB photons which converts $E$-modes into $B$-modes \cite{zaldarriaga_seljak_1998} and therefore their detection -- a source of constraints on the matter perturbation evolution at redshift $z\sim1$ when light massive neutrinos and elusive dark energy both play potentially visible roles. 

 For these reasons, many polarization experiments targeting $B$-modes have been built or proposed, including  ground based observatories, those already operating {\it e.g.}, \textsc{polarbear} \cite{polarbear} or \textsc{sptpol}~\cite{sptpol}  or those, which are being developed, {\it e.g.}, {\sc qubic} \cite{qubic}, \textsc{actpol} \cite{actpol}, balloon borne experiments such as {\sc spider} \cite{spider} or \textsc{ebex} \cite{ebex}, which flew in winter 2012/13,  or to even a potential satellite mission {\sc l}ite\textsc{bird}~\cite{litebird}, \textsc{co}r\textsc{e}\cite{bpol_website}, \textsc{pix}i\textsc{e}~\cite{pixie}. 
 With an exception of the \textsc{qubic} experiment, all these experiments scan the sky with one or more dishes and therefore most directly produce maps of the polarized Stokes parameters, $Q$ and $U$.
 Calculation of the $E$ and $B$ signals from the $Q$ and $U$ maps is a non-local~operation \cite{zaldarriaga_2001} and can be done uniquely only if the full sky maps are available. However, this can be hardly the case even for the satellite missions due to the presence of heavy non-cosmological contamination due to the Galactic emissions, which typically have to be masked out even after advanced and complex cleaning procedures have been applied.
In the context of the pseudo-spectrum methods \cite{hauser_peebles_1973, hansen_gorski_2003, hivon_etal_2002} the incomplete sky coverage leads to the so called $E$-to-$B$ leakage, when the signal from $E$-modes is present in the reconstruction of $B$-modes power spectrum $C_{\ell}^{B}$ and more problematic, in the $B$-modes uncertainties. Though no bias is  directly introduced, the leakage is a problem due to the much higher amplitudes of the $E$-modes signal, which then 
 inflates the overall  uncertainty of the  estimated $B$-modes signal potentially precluding its detection. 

Several extensions of the standard pseudo-spectrum methods have been recently proposed designed to alleviate the $E$-to-$B$ leakage problem. In this work we focus on the technique presented in Refs. \cite{smith_2006,smith_zaldarriaga_2007,grain_etal_2009}  and working in the harmonic domain and referred to as the \sz-method hereafter, and on two other techniques operating in the pixel domain presented in Ref. \cite{zhao_baskaran_2010} and Refs. \cite{kim_naselsky_2010,kim_2010}, and referred to as the \zb- and \kn-techniques\footnote{The methods' names are based on the first letters of the names of the authors of the corresponding papers.}, respectively. All these methods consist in filtering $E$-modes leaking into $B$-modes for each specific realization of the polarized anisotropies and thus potentially resolve the excessive variance problem referred to earlier.

In this article, we first describe each of these methods within a common framework of so-called $\chi$-fields and then our implementations of them, emphasizing differences and similarities with those proposed in the original 
papers. 
Throughout this work we compute spatial derivatives of the sky maps in the harmonic domain.  This is in agreement with the original implementations of the considered techniques.
We note however that an interesting, pixel-domain alternative has been recently proposed in Ref.~\cite{bowyer_2011} and could be exploited in future work.
For spectrum estimators we use consistently cross-spectra \cite{tristram_etal_2005},  rather than auto-spectra, avoiding therefore a need for estimating the instrumental noise spectrum. 

We use numerical experiments to test the efficiency of each of these methods in terms of quality of the $C_{\ell}^{B}$ reconstruction and above all of the resulting uncertainty.  
The numerical experiments involve two experimental set-ups: one mimicking a satellite mission (loosely based on {\sc epic} ~\cite{bock_etal_2008}), and, the other,  a balloon-borne instrument (inspired by {\sc ebex} \cite{britt_etal_2010}).  We note that this kind of analyses of satellite-mission-like set-ups are largely absent  in the literature, which predominantly has focused on small-sky cases only. Though other techniques, e.g., maximum-likelihood based power spectrum estimators, may address better some of the problems faced by nearly-full sky observations, performance of the pseudo-spectrum methods in this regime is clearly of practical importance.

The general pseudo-spectrum formalism, as well as its standard and extended renditions relevant for this work, are introduced in section \ref{sect:polarization}.  An overview of the methods and their implementations can be found in section \ref{sect:formalism}. The numerical results are given in section \ref{sect:results}, which also presents the case for  the \sz-method as the one which gives the smallest variances while avoiding a bias.
More extensive conclusions are then given in section \ref{sect:conclusion}, while technical details are deferred to appendices, with App.  \ref{app:noise} treating the problem of the noise bias for the \zb~and \kn~methods.


\section{Pseudo-spectrum polarized power spectrum estimators}
\label{sect:polarization}
\subsection{General considerations}
The linearly polarized CMB polarization field is completely described by a spin-(2) and a spin-(-2) fields, $P_{\pm2}(\vec{n})=Q(\vec{n})\pm iU(\vec{n})$, with $Q$ and $U$ denoting two Stokes parameters. Pseudo-spectrum methods distill the observed information into a set of harmonic coefficients, $\tilde{a}^E_{\ell m}$ and $\tilde{a}^B_{\ell m}$, referred to as {\em pseudo-}multipoles. These are related to {\em true} multipoles,
$a^E_{\ell m}$ and $a^B_{\ell m}$ as follows,
\begin{eqnarray}
	\tilde{a}^E_{\ell m}&=&\ds\sum_{\ell'm'}\left[H^{(+)}_{\ell m,\ell'm'}a^E_{\ell'm'}+iH^{(-)}_{\ell m,\ell'm'}a^B_{\ell'm'}\right],
	\label{eq:aEpseudoGenDef}
	\\
	\tilde{a}^B_{\ell m}&=&\ds\sum_{\ell'm'}\left[-iK^{(-)}_{\ell m,\ell'm'}a^E_{\ell'm'}+K^{(+)}_{\ell m,\ell'm'}a^B_{\ell'm'}\right],
	\label{eq:aBpseudoGenDef}	
\end{eqnarray}
where $H^{(\pm)}$ and $K^{(\pm)}$ are kernels, which in general can be all different, non-vanishing, and non-diagonal in both $\ell$ and $m$.
Noise terms have been neglected in these equations for shortness.

The kernels  are typically singular and it is not in general possible to solve the inverse problem to recover the true multipoles, $a_{\ell m}^X$, directly. 
Instead the pseudo-spectrum approaches attempt to do so only on the power spectrum level. This is achieved in two steps.
First, owing to the statistical isotropy of CMB fluctuations, we can rewrite Eqs.~\eref{eq:aEpseudoGenDef}  and~\eref{eq:aBpseudoGenDef}  on the power spectrum level as,
\begin{eqnarray}
	\left<\tilde{C}^E_{\ell}\right> &=&\ds\sum_{\ell'}\left[H^{(+)}_{\ell\ell'}\left<C^E_{\ell'}\right>+H^{(-)}_{\ell\ell'}\left<C^B_{\ell'}\right>\right],
	\label{eq:cEpseudoGenDef}\\
	\left<\tilde{C}^B_{\ell}\right> &=&\ds\sum_{\ell'}\left[K^{(-)}_{\ell\ell'}\left<C^E_{\ell'}\right>+K^{(+)}_{\ell\ell'}\left<C^B_{\ell'}\right>\right].
	\label{eq:cBpseudoGenDef}	
\end{eqnarray}
where the new kernels,  $X^{(\pm)}_{\ell\ell'}$ are given by ($X=K, H$),
\begin{equation}
	X^{(\pm)}_{\ell\ell'}=\ds\sum_{m'=-\ell'}^{\ell'}\frac{1}{2\ell+1}\ds\sum_{m=-\ell}^{\ell}\left|X^{(\pm)}_{\ell m,\ell'm'}\right|^2,
	\label{eq:psKernelDef}
\end{equation}
and $\left<\dots\right>$ denotes an ensemble average and,
\begin{equation}
\tilde{C}_\ell^X \equiv \frac{1}{2\ell+1}\,\sum_{m=-\ell}^\ell\,\left|\tilde{a}_{\ell m}^X\right|^2.
\label{eq:psudoSpecDef}
\end{equation}
The kernels obtained on the power spectrum level are clearly more manageable and easier to calculate, nevertheless, they still will be 
singular. To avoid this issue, the inverse problem defined in Eqs.~\eref{eq:cEpseudoGenDef} and~\eref{eq:cBpseudoGenDef} is solved only
for binned spectra~\cite{hivon_etal_2002}, 
\begin{equation}
\label{eq:binnedSpec}
\begin{array}{l c l}
{\displaystyle \tilde{C}_b^X} &{\displaystyle \equiv}& {\displaystyle \sum_\ell \; P_{b\ell}\, \tilde{C}^X_\ell}
\\
{\displaystyle C_b^X}& {\displaystyle \equiv} & {\displaystyle \sum_\ell \; P_{b\ell}\,C_\ell^X,}
\end{array}
\end{equation}
where the binning operators are defined as,
\begin{eqnarray}
	P_{b\ell}&=&\left\{\begin{array}{cl}
		{\displaystyle \frac{S_{\ell}}{\ell^b_\mathrm{max}-\ell^b_\mathrm{min}},} & {\displaystyle \ell\in[\ell^b_\mathrm{min},\ell^b_\mathrm{max}]} \\
		{\displaystyle 0,} &  {\displaystyle  \ell\notin[\ell^b_\mathrm{min},\ell^b_\mathrm{max}]}
	\end{array}\right. \nonumber \\
	Q_{b\ell}&=&\left\{\begin{array}{cl}
		{\displaystyle  \ \ \  \ \ \ \frac{1}{S_{\ell}},\ \ \ \ \ \ } & {\displaystyle  \ell\in[\ell^b_\mathrm{min},\ell^b_\mathrm{max}] }\\
		{\displaystyle 0,} & {\displaystyle  \ell\notin[\ell^b_\mathrm{min},\ell^b_\mathrm{max}]},
	\end{array}\right. \nonumber
\end{eqnarray}
satisfying therefore the relation $\sum_{\ell} Q_{b\ell} \, P_{b'\ell} = \delta_{bb'}$. Here, we have introduced a shape function, $S_\ell$. Its role is to minimize possible binning effects 
by making $S_\ell\,\tilde{C}$ nearly flat within the bin. Hereafter, we will adopt the standard choice for it, {\it i.e.}, $S_\ell = \ell(\ell+1)/2\pi$.
The binned version of Eqs.~\eref{eq:cEpseudoGenDef} and~\eref{eq:cBpseudoGenDef}  now reads,
\begin{equation}
\label{eq:binnedSystem}
	\left(\begin{array}{c}
		\tilde{C}^E_b \\
		\tilde{C}^B_b
	\end{array}\right)\simeq\displaystyle\sum_{b'}\left(\begin{array}{cc}
		H^{(+)}_{bb'} & H^{(-)}_{bb'} \\
		K^{(-))}_{bb'} & K^{(+)}_{bb'}
	\end{array}\right)\left(\begin{array}{c}
		{C}^E_{b'} \\
		{C}^B_{b'}
	\end{array}\right),
\end{equation}
where, for $X = K$ or $H$,
\begin{equation}
X_{bb'} \equiv \sum_{\ell,\ell'}P_{b\ell}X_{\ell\ell'}Q_{b'\ell'}.
\label{eq:binnedKernelDef}
\end{equation}
To include a correction for the presence of the instrumental noise, the pseudo-power spectrum on the right hand side of the first of Eqs.~\eref{eq:binnedSpec} needs be corrected for the noise pseudo
spectrum prior to the binning operations.

The estimates of the true spectra, $C_\ell^X$, can be then obtained by directly solving the full system in Eq.~\eref{eq:binnedSystem}. We note that by construction, and neglecting the binning effects, which are largely controllable, these will be unbiased estimates of the true binned spectra.
However, as long as the polarization mode mixing kernel, $K^{(-)}$, does not vanish\footnote{Strictly speaking what is required is that the multipole kernel, $K^{(-)}_{\ell m, \ell' m'}$ vanishes but if Eq.~\eref{eq:psKernelDef} is 
satisfied, exactly or approximately, it is equivalent to requiring the power spectrum kernel, $K^{(-)}_{\ell\ell'}$, to be (nearly) zero.} the power contained in the $E$-polarization component will contribute to the overall variance of the $B$-spectrum estimate
-- an effect referred to as the $E$-to-$B$ leakage. To avoid that one should resort to methods for which $K^{(-)}$ is either zero or nearly so. We also note that if $K^{(-)}= 0$ then the estimate 
of the $B$-mode spectrum can be derived independently on the $E$ one. This could be also the method of choice even if $K^{(-)}$ vanishes only approximately. In this  case a small bias in the $B$ spectrum
estimate is however to be expected.

\subsection{Standard pseudo spectrum approach}

If the polarization fields are known on the entire celestial sphere, their $E$- and $B$-representation can be easily obtained in the harmonic domain using the spin-weighted spherical harmonics\footnote{All the integrals in this paper are taken over the {\it entire} celestial sphere. We therefore do not specify that the integration domain is $S^2$.},
\begin{eqnarray}
\begin{array}{l c l}
\medskip
	a^E_{\ell m}&=&{\displaystyle \frac{-1}{2}}\ds\int \left[P_{2}(\vec{n})\,{}_{2}Y^\star_{\ell m}(\vec{n})+P_{-2}(\vec{n})\,{}_{-2}Y^\star_{\ell m}(\vec{n})\right]d\vec{n}, \\
 	a^B_{\ell m}&=&{\displaystyle \frac{i}{2}}\ds\int \left[P_{2}(\vec{n})\,{}_{2}Y^\star_{\ell m}(\vec{n})-P_{-2}(\vec{n})\,{}_{-2}Y^\star_{\ell m}(\vec{n})\right]d\vec{n}.
\end{array}
\end{eqnarray}
If the polarization field is measured on a fraction of the sky only, the above decomposition can be most straightforwardly applied to such a case by positing that the signal over the
unobserved part of the sky vanishes..
This choice defines the standard pseudo-spectrum method, in which the resulting {\em pseudo-}multipoles, $\tilde{a}^X_{\ell m}$, $X = E, B$, can be expressed as follows,
\begin{eqnarray}
	\tilde{a}^E_{\ell m}&\equiv&\frac{-1}{2}\ds\int M\left[P_{2}(\vec{n})\,{}_{2}Y^\star_{\ell m}(\vec{n})+P_{-2}(\vec{n})\,{}_{-2}Y^\star_{\ell m}(\vec{n})\right]d\vec{n} \nonumber \\
	&=&\ds\sum_{\ell'm'}\left[K^{(+)}_{\ell m,\ell'm'}a^E_{\ell'm'}+iK^{(-)}_{\ell m,\ell'm'}a^B_{\ell'm'}\right],
	\label{eq:stdElm}
\\
	\tilde{a}^B_{\ell m}&\equiv&\frac{i}{2}\ds\int M\left[P_{2}(\vec{n})\,{}_{2}Y^\star_{\ell m}(\vec{n})-P_{-2}(\vec{n})\,{}_{-2}Y^\star_{\ell m}(\vec{n})\right]d\vec{n} \nonumber \\
	&=&\ds\sum_{\ell'm'}\left[-iK^{(-)}_{\ell m,\ell'm'}\,a^E_{\ell'm'}+K^{(+)}_{\ell m,\ell'm'}\,a^B_{\ell'm'}\right],
	\label{eq:stdBlm}
\end{eqnarray}
where $M$ is a binary mask defining observed patch, and where we introduced the convolution kernels, $K^{(\pm)}_{\ell m,\ell'm'}$, explicit expressions for which are well-known and can be found 
elsewhere, {\it e.g.}, ~\cite{grain_etal_2009}.
We see that  for the standard technique both the $H^{(\pm)}$ and $K^{(\pm)}$ kernels, Eqs.~\eref{eq:aEpseudoGenDef}~\&~\eref{eq:aBpseudoGenDef}, coincide and that the polarization-mode 
mixing kernel, $K^{(-)}_{\ell m, \ell' m'}$, does not vanish and therefore
though unbiased, the standard pseudo-power spectrum estimator suffers from the $E$-to-$B$ leakage. This can be quite severe. For instance, an experiment covering around $1$\% of the sky 
essentially unable to detect a power at the scales larger than $\ell \simlt 140$ (see Fig. 16 of \cite{grain_etal_2009}).

The above formulae can be extended to include an arbitrary weighting of the observed sky pixels as given by a window function, $W$. This can be done by inserting $W \, M$ instead of $M$ in all the
equations above, including those for the kernels. If  we further assume that the window function
is always zero outside of the observed sky, {\it i.e.} if $M = 0$ then also $W=0$, then, as a consequence, $W \, M  =  W$ and $M$ can be dropped from the equations in favor of $W$. The mask, $M$, is then
assumed to be defined implicitly by $W$. We will use this simplification in the following.
Also for definiteness hereafter, we assume that
a field defined on the sphere, {\it e.g.}, $P_{\pm 2}$,  is known  on the full sky and will apply a mask or an apodization explicitly  to such a field to emphasize that it is known only over a limited
sky area, e.g., $W\,P_{\pm 2}$.

\subsection{Leakage-free pseudo-power spectrum approaches}

To alleviate the leakage problem within the pseudo-spectrum methods one would need to adapt a different definition of the pseudo-multipoles than the one used in the standard approach.
Such a new definition should not rely directly on the polarization fields, as does the standard
approach, as those unavoidably incorporate contributions from both types of polarized multipoles.
 Instead it should based on some other fields, which depend only on one set of the multipole coefficients, and which would therefore ensure that the polarisation
mode mixing kernels, $K^{(-)}_{\ell m, \ell' m'}$ and $H^{(-)}_{\ell m, \ell' m'}$, indeed vanish, resolving the leakage issue.

Such a construction has been indeed proposed by~\cite{zaldarriaga_seljak_1997}
and the corresponding fields are called $\chi$-fields. They can be derived from the polarization fields as follows,
\begin{eqnarray}
	\chi^E(\vec{n})&=&-\frac{1}{2}\left[\bar\partial\bar\partial P_{2}(\vec{n})+\partial\partial P_{-2}(\vec{n})\right],
	\label{eq:chiEdef}
	 \\
	\chi^B(\vec{n})&=&\frac{i}{2}\left[\bar\partial\bar\partial P_{2}(\vec{n})-\partial\partial P_{-2}(\vec{n})\right],
          \label{eq:chiBdef}
\end{eqnarray}
where $\partial(\bar\partial)$ denotes the spin-raising(lowering) operator~\cite{zaldarriaga_seljak_1997}.
These $\chi$ fields involve indeed either $E$-, in the case of  $\chi^E$, or $B$-, for $\chi^B$, modes.  This can be seen directly by noting that the $\chi^X$-fields, $X=E, B$, are scalar and given by,
\begin{equation}
	\chi^{X}(\vec{n}) \, = \, \sum_{\ell,m}\,N_{\ell, 2}\, a^{X}_{\ell m}\,Y_{\ell m}(\vec{n}),
	\label{eq:chiPure}
\end{equation}
where for the future convenience we have introduced,
\begin{equation}
N_{\ell, s}\equiv\sqrt{\frac{(\ell+s)!}{(\ell-s)!}}. \nonumber
\end{equation} 
In the full-sky case, Eq.~\eref{eq:chiPure} can be readily inverted giving,
\begin{equation}
	\chi^X_{\ell m}=\ds\int \chi^{X}(\vec{n})Y^\star_{\ell m}(\vec{n}\,)d\vec{n}  = N_{\ell, 2}\, a^{X}_{\ell m},
	\label{eq:chiDecompFullSky}
\end{equation}
what in turn can be  adapted for cases of partial sky experiments in a usual manner, rendering the following 
definition of the pseudo multipoles,
\begin{equation}
\tilde{a}^{X}_{\ell m} \equiv  \frac{1}{N_{\ell, 2}}\, \ds\int M(\vec{n})\,\chi^{X}(\vec{n})Y^\star_{\ell m}(\vec{n})\,d\vec{n}.
\label{eq:pseudoAlmViaChi}
\end{equation}
This definition can be then used in the general pseudo-spectrum formalism as developed in Sect.~\ref{sect:polarization} and though it will result in a mixing of different $\ell$-modes, it will not cause any leakage between 
the polarization modes as by construction the off-diagonal kernels, ${H}^{(-)}$ and ${ K}^{(-)}$ in Eqs.~\eref{eq:aEpseudoGenDef} and~\eref{eq:aBpseudoGenDef}, vanish.

The major difficulty of this approach is the computation of the $\chi^X$-fields. Indeed, Eqs.~\eref{eq:chiEdef} \&~\eref{eq:chiBdef}, as they are, require in principle knowledge of the full sky polarization fields.
As we will see in the next section all three methods designed to resolve the leakage problem and studied in this work rely on the $\chi^X$ field calculation,  implicitly or explicitly, and circumvent the problem of having only a limited sky coverage differently.

We note that if the $\chi^X$ fields were known exactly on the cut sky, the inverse problem in Eq.~\eref{eq:binnedSystem}, could be solved separately for $E$ and $B$ spectra, as the off-diagonal kernels would, by construction, vanish. In more realistic circumstances the $\chi^X$ fields, actually estimated  on the cut sky, may be imperfect giving, at least in principle, rise to non-zero off-diagonal contributions. These, if not corrected for, could lead to a bias of
the estimated power spectra. Solving the full system, accounting for the non-diagonal kernels, could help to trade the bias for an extra, but presumably small variance of the spectrum estimate. 
Though this indeed could be possible at least for some of the methods, for others, the difficulty in calculating the off-diagonal kernels, either analytically or numerically, {\it e.g.}, via Monte Carlo simulations, can be prohibitive, and an approach favored in practice is often simply to accept the bias, once it is found to be sufficiently small. 

\section{Specific approaches}

\label{sect:formalism}


\subsection{\sz-approach}
\label{subsec:pure}
\subsubsection{Theoretical description} 
Let us start from the pseudo-multipoles for $B$-modes defined as in Eq.~\eref{eq:pseudoAlmViaChi} with the binary mask, $M$, replaced by an arbitrary window, $W$.
 By performing an integration by parts twice~\cite{smith_2006, smith_zaldarriaga_2007}, we can rewrite this equation as,
\begin{eqnarray}
\label{eq:aPureDef}
	\tilde{a}^B_{\ell m}&=&\frac{i}{2N_{\ell,2}}\ds\int d\vec{n}\bigg[P_{2}(\vec{n})\times\left(\partial\partial W(\vec{n})Y_{\ell m}(\vec{n})\right)^\star  \\
	&-&P_{-2}(\vec{n})\times\left(\bar\partial\bar\partial W(\vec{n})Y_{\ell m}(\vec{n})\right)^\star\bigg], \nonumber
\end{eqnarray}
where all the boundary terms are omitted corresponding to an assumption that  the apodization window, $W(\vec{n})$ and its first derivative, $\partial W$, vanish at the observed patch boundaries. This latter equation has an
advantage over the former, Eq.~\eref{eq:pseudoAlmViaChi}, as it does not involve any explicit calculation of derivatives of noisy sky maps. Instead, the differentiation needs to be only applied to a presumably smooth window function, $W$. We can therefore
use Eq.~\eref{eq:aPureDef} as a definition of the pseudo-multipoles, which we will apply from now on also in cases when the apodization does not conform with the boundary conditions.
Note that in these latter cases there will be no assurance that no $E$-to-$B$ leakage is present. 

Hereafter we will refer to this technique as a pure pseudo-spectrum estimator, as Eq.~\eref{eq:aPureDef} can be interpreted as projecting the
polarization field $P_{\pm 2}$ onto a basis of 'pure' functions representing only $B$-like polarization modes on a cut sky~\cite{bunn_etal_2003,smith_2006, smith_zaldarriaga_2007}.

\subsubsection{Numerical implementation} 
Our implementation of the approach follows closely that proposed in \cite{grain_etal_2009} and proceeds in four steps.

\medskip
\noindent$\bullet$~{\it Step 1}: 
\medskip

We compute spin-0, spin-1 and spin-2 renditions of the window function, $W$, given by,
\begin{equation}
	W_0=W,~W_1=\partial W,~W_2=\partial\partial W.
\end{equation}
Because $W$ is real, then  $W^\star_s=W_{-s}$ for a spin $s=1, 2$.

\medskip
\noindent$\bullet$~{\it Step 2}: 
\medskip

We compute pure pseudo-multipoles  by constructing first three apodized maps,
\begin{equation}
	\mathcal{P}_{\pm2}=W_0P_{\pm2},~\mathcal{P}_{\pm1}=W_{\mp1}P_{\pm2},~\mathcal{P}_{\pm0}=W_{\mp2}P_{\pm2}.
\end{equation}
and then calculating pure $\tilde{a}^B_{\ell m}$ as,
\begin{equation}
	\tilde{a}^B_{\ell m}=\frac{1}{N_{\ell,2}}\left(\mathcal{B}_{0,\ell m}+2N_{\ell,1}\mathcal{B}_{1,\ell m}+N_{\ell,2}\mathcal{B}_{2,\ell m}\right),
	\label{equ:purBpract}
\end{equation}
where $\mathcal{B}_{s,\ell m}$ is a $B$-type mutlipole of $\mathcal{P}_{\pm s}$ defined as
\begin{eqnarray}
	\mathcal{B}_{s,\ell m}&=&\frac{i}{2}\ds\int \Big[\mathcal{P}_{+s}(\vec{n}){}_{s}Y^\star_{\ell m}(\vec{n}) \\
	&&-(-1)^s\mathcal{P}_{-s}(\vec{n}){}_{-s}Y^\star_{\ell m}(\vec{n})\Big]d\vec{n}. \nonumber
\end{eqnarray}

\medskip
\noindent$\bullet$~{\it Step 3}: 
\medskip

On this step we compute the convolution kernels for pseudo-$C_\ell$ as defined in Eqs.~\eref{eq:cEpseudoGenDef} \&~\eref{eq:cBpseudoGenDef}.
This can be done using, {\it e.g.}, Eqs. (A13) and (A14) of~\cite{grain_etal_2009}. If the applied apodization does not fulfill the boundary conditions then the off-diagonal block, $K^{(-)}$, 
has to be also included. In practice, the off-diagonal coupling between the polarization components will also result  due to pixelization effects. Though such effects are not accounted for
in the analytic formulae for the kernels, they can be corrected for, to some extent,  by a procedure described in~\cite{grain_etal_2009},
leading to a removal of the majority of small bias induced by the residual, pixel-induced, $E$-to-$B$ leakage.

We note that typically, if the method is applied consistently to both $E$ and $B$-modes the corresponding $H$ and $K$ kernels are identical. However, in some circumstances it may be advantageous and possible to apply hybrid
approaches in which both kinds of spectra are treated differently. Such cases have been discussed recently in~\cite{grain_etal_2012}.

\medskip
\noindent$\bullet$~{\it Step 4}:
\medskip

This step consists of standard operations involved in any pseudo-spectrum method as summarized by Eqs.~\eref{eq:binnedSpec} \&~\eref{eq:binnedSystem} and discussed in Sect.~\ref{sect:polarization}.

\subsubsection{Sky apodization}
As emphasized  in \cite{smith_2006,smith_zaldarriaga_2007,grain_etal_2009}, an appropriate sky apodization is a key element of any such a construction. In the specific method discussed here the degree to which the apodization fulfills the boundary conditions will be a principal factor
determining the level of a suppression of the $E$-to-$B$ leakage. At the same any apodization applied to realistic, meaning noisy, data will have a direct impact on the resulting uncertainties of the spectrum estimate.
In the context of the pure pseudo-spectrum method,  
 systematic approaches have been developed and studied in detail, which allow for a numerical optimisation of  sky apodizations in order to ensure a nearly minimal value of the final spectrum uncertainty~\cite{smith_2006, smith_zaldarriaga_2007, grain_etal_2009}.  These are either based on MC simulations or semi-analytic techniques.
In the former case, MC simulations are used to tune the apodization length of the sky apodization given by some analytic formul\ae. In this work, we will use the so-called $C^2$ function as given by equation (31) of \cite{grain_etal_2009}. In the latter case, the optimized sky apodization can be computed by solving a large linear system as proposed in~\cite{smith_zaldarriaga_2007}. We refer to these latter windows as 
 {\it variance-optimized apodization}.
In both cases the optimization could, and 
should,  be applied bin-by-bin to ensure the best results. As discussed at length in~\cite{grain_etal_2009} both these approaches require some prior assumptions concerning, for instance, the angular power spectra of 
$E$- and $B$-modes, however, the results of the optimisation are found to be only mildly dependent on details of the assumed $B$-mode spectrum.

It has been shown via numerical experiments \cite{grain_etal_2009} that the variance-optimized apodizations lead systematically to the lowest error bars on the reconstructed $C^B_\ell$'s and therefore will be used them in this work. Those variance-optimized apodizations can be computed in two ways, depending on the domain (harmonic domain or pixel domain) in which the linear system is solved. For the peculiar case of homogeneous noise, resolution can be vastly done in the harmonic domain. In such a case, the derivative relationship $W_{s=1,2}=\partial^s W_0$ and the boundary conditions $W_0(\mathcal{C})=W_1(\mathcal{C})=0$ on the contour of the observed region are fulfilled (up to pixelization effects). For more general cases, the linear system providing the variance-optimized apodization is solved in the pixel domain. In such a setting, both the derivative relationship and the boundary conditions are relaxed ($W_0,~W_1$ and $W_2$ are considered as independent). As a consequence, the final sky apodizations does not strictly satisfy these conditions and the resulting pseudo-multipoles will not be strictly equal to the pure pseudo-multipoles.
However, it has been shown in \cite{smith_zaldarriaga_2007,grain_etal_2009} that the angular power spectra recovered in such cases consistently achieve smaller uncertainties than those of other
apodization choices.

\subsection{\zb-approach}
\label{subsec:zb}
\subsubsection{Theoretical description}
In this approach the $\chi^{X}$ fields are computed directly in the pixel domain and for the cut-sky. 
This is made possible thanks to a formula derived in \cite{zhao_baskaran_2010}, which reads,
\begin{eqnarray}
	W(\vec{n})\chi^B(\vec{n})&=&\frac{i}{2}\left[\bar\partial\bar\partial\left(WP_{2}\right)-\partial\partial\left(WP_{-2}\right)\right] 
	\label{eq:chiB}
	\\
	&-&i\left[\frac{\bar\partial W}{W}\bar\partial\left(WP_{2}\right)-\frac{\partial W}{W}\partial\left(WP_{-2}\right)\right] \nonumber \\
	&-&\frac{i}{2}\left[\left(\bar\partial\bar\partial W\right)P_{2}-\left(\partial\partial W\right)P_{-2}\right]. \nonumber \\
	&+&i\left[\frac{\left(\bar\partial W\right)^2}{W}P_{2}-\frac{\left(\partial W\right)^2}{W}P_{-2}\right]. \nonumber
\end{eqnarray}
As usual here $W$ is assumed to be zero outside the observed region. Moreover, if we assume that it and its first derivative vanish at the edges of the observed region, all the
operations on the right hand side of this equation can be performed with only knowledge of the polarization field  on the cut-sky.
Consequently, we could estimate the field, $\chi^B$ consistently on the cut-sky by first computing the rhs of Eq.~\eref{eq:chiB}, then dividing it  by the window, $W$, and later use it to calculate pseudo-multipoles via Eq.~\eref{eq:pseudoAlmViaChi} -- as proposed in \cite{zhao_baskaran_2010} -- or use some apodized rendition of the $\chi^B$ field to derive the pseudo-multipoles, which are then corrected on the
power spectrum level -- as proposed here\footnote{Strictly speaking, the pseudo-multipole are not divided by $N_{\ell,2}$ in the implementation of Ref. \cite{zhao_baskaran_2010}. Instead, the pseudo-spectrum are divided by $N^2_{\ell,2}$ in the binning process. The two choices are however completely equivalent.}.
In either case the pseudo-multipoles are in principle free of any $E$-to-$B$ leakage due to cut sky effects and the $K^{(-)}_{\ell\ell'}$ kernel should vanish. However, as underlined by \cite{zhao_baskaran_2010}, both pixelization and convolution by the beam lead to some residual $E$-to-$B$ leakage and ideally one would like to solve the full linear system, Eq.~\eref{eq:binnedSystem}, to get the final, unbiased power spectrum estimation.

\subsubsection{Numerical implementation}
An implementation of this technique is proposed in \cite{zhao_baskaran_2010} and involves four steps.  The implementation used in this work follows that of the original authors with an exception of the second step as detailed below.

\medskip
\noindent$\bullet$~{\it Step 1}:
\medskip
 
We compute the $\chi^B$ field on the observed patch of the sky using Eq. (\ref{eq:chiB}). This in turn requires a numerical calculation of derivatives of noisy fields, which constitutes the principal
 difficulty of this technique.  These in our implementation, as well as that of~\cite{zhao_baskaran_2010} are performed in the harmonic domain. We emphasize that with such a choice this method becomes effectively a harmonic space approach. Yet another potential problem is related to the calculation of the terms, which involve explicit multiplication by $W^{-1}\,\partial\,W$, as $W$ itself becomes very small at the boundary. This problem cannot be avoided by imposing more boundary conditions on $W$ as  $W^{-1}\partial W \sim\left|\theta-\theta_c\right|^{-1}$,  at the boundary, $\theta_c$, and therefore necessarily diverges at the boundary\footnote{Constraining $W$ together with its first derivative $\partial W$, both to be continuous on the entire celestial sphere but zero outside the observed part of the sky necessary leads to $W\sim\left|\theta-\theta_c\right|^{n}$ with $n\geq1$, close to the boundary.}. 
 This can be however dealt with on Step 2.
 
\medskip
\noindent$\bullet$~{\it Step 2}:
\medskip

 We compute the pseudo-multipoles, $\tilde{a}^{B}_{\ell m}$, of the newly constructed $\chi^B$ map. This requires effectively dividing by the window, $W$. Though straightforward a priori a care has to be exercised while doing so  because of $W$ vanishing at the observed area edges.

One option, adopted in \cite{zhao_baskaran_2010}, relies on simple trimming the troublesome, boundary layer, leaving only those pixels for which the division is numerically reliable. This leads to some loss of 
the information but solves simultaneously the divergence problem appearing on step 1. The amount  lost  due to trimming will depend on the details of how the trimming is done, a practical complication, which needs to be addressed
in this approach.

An alternative way of resolving both these issues at the same time, which we propose here and which is free of such practical complications, is to define pseudo-multipoles using the field, $W^2\,\chi^B$, and then to correct for the presence of the apodization on the binned spectrum estimation step, Eq.~\eref{eq:binnedSystem}. It is clear from Eq.~\eref{eq:chiB} that the estimation of the $W^2\,\chi^B$ field does not suffer of any singularities at the edges. This method is the method of choice in this work.

We note that this method is not lossless either, as the apodization it invokes  will unavoidably compromise some information. Nevertheless, the information loss in this case  is expected to be
smaller than in the former one. For instance,
it is argued in Sec.~IV of \cite{zhao_baskaran_2010} that to analyze a map covering 3\% of the sky (a spherical cap with a radius of 20 degrees is assumed as the observed part of the sky), it is necessary to remove an external layer with a width of 2 degrees; thus reducing the effective sky coverage from 3\% to 2.4\% (assuming a binary mask to weight the resulting $\chi^B$ map). As shown hereafter, by focusing on $W^2\,\chi^B$, we are able to solve for the $E$-to-$B$ leakage by using an apodization length of 1 degree. As a consequence, for a spherical cap with a radius of 20 degrees, the effective sky coverage is reduced from 3\% to 2.9\% (an explicit expression for the effective sky coverage assuming non-binary mask can be found in \cite{grain_etal_2012}).

\medskip
\noindent$\bullet$~{\it Step 3}:
\medskip

Kernel $K^{(+)}_{\ell\ell'}$ is computed taking advantage of the fact that the $\chi^B$ field is a scalar, like temperature, made of $B$-modes. The explicit expression of $K^{(+)}_{\ell\ell'}$ is given by Eq. (39) of \cite{zhao_baskaran_2010} (following what was derived for temperature \cite{hivon_etal_2002,tristram_etal_2005,hauser_peebles_1973,hinshaw_etal_2003}), i.e.,
\begin{equation}
\label{eq:kernelScalar}
	K^{(+)}_{\ell\ell'}=\frac{(2\ell'+1)N^2_{\ell',2}}{4\pi N^2_{\ell,2}}\ds\sum_{\ell''m''}\left|w^{(2)}_{\ell''m''}\right|^2\left(\begin{array}{ccc}
		\ell & \ell' & \ell'' \\
		0 & 0 & 0
	\end{array}\right)^2,
\end{equation}
with $w^{(2)}_{\ell''m''}$ the multipoles of the $W^2$ function\footnote{We stress that the multipoles of $W^2$ are {\it not} equal to the square of the multipoles of $W$.}.

\medskip
\noindent$\bullet$~{\it Step 4}: 
\medskip

The linear system in Eq. (\ref{eq:binnedSystem}) is inverted neglecting the off-diagonal block, $K^{(-)}_{\ell\ell'}$, and therefore also the residual $E$-to-$B$ leakage.

\subsubsection{Sky apodization}
In this approach we could either use analytic windows or the variance-optimized windows obtained from the optimization procedure developed within the framework of the 
sz-method. In this former case, we will always use the $C^2$ family of windows from  Ref. \cite{grain_etal_2009} and use MC simulations to determine their optimal apodization length.

In the case of the  variance-optimized apodizations computed in the harmonic domain, it may appear that to ensure their optimality, we should use a window given by  a square root 
of the actual optimized one, i.e., $W_{ZB} \equiv \sqrt{W_{SZ}}$,  to compensate for the fact that it is a square of the window which is used as the apodization  in our implementation 
of the \zb-approach.
Whether such a window could be a viable option, will depend whether it does not cause any problems in the calculation of the rhs of Eq.~\eref{eq:chiB} at the patch edges. 
It is straightforward to show that this is always the case for windows, which are forced to obey the boundary conditions strictly.
This is because such windows scale at the boundary as $W_{SZ}\sim \left|\theta-\theta_c\right|^n$, with $n>2$,~\cite{smith_zaldarriaga_2007} therefore both quantities, $\partial W_{ZB}$ and $\partial\partial W_{ZB}$, (where $W_{ZB} = \sqrt{W_{SZ}}$), needed to compute the rhs of  Eq.~\eref{eq:chiB} are well-behaved for $\theta \sim \theta_c$. 
However, the variance-optimized windows fulfil the boundary condition only approximately, what may lead to singularities of the derivatives of $W_{ZB}$. To avoid that, we further multiply the variance-optimised windows by some analytic window, with a narrow apodizaton length. This is designed to affect as little as possible the properties of the initial window but enforce the boundary conditions strictly and therefore
ensure proper behaviour of the resulting window at the boundary. In practice, we have found that using either the corrected $W_{ZB}$ window  or directly $W_{SZ}$ leads to comparable results and numerical results presented
hereafter are using the latter ones.

It is important to notice that in such settings, the variance-optimized windows computed in the pixel domain cannot be directly applied. Indeed, such windows do not conform typically with  the derivative relationship between the different windows, i.e., $W_{s=1,2}\neq \partial^s W_0$ or the boundary conditions, i.e.,  $W_0(\mathcal{C}) \neq W_1(\mathcal{C})=0$. However, these conditions are essentially mandatory for the {\sc zb}-method for two reasons. First, the method requires that $W\,\chi^B$ is related to  $W_{s=0,1,2}\, P_{\pm2}$ and $M\,P_{\pm2}$, as e.g., it is in Eq.~\eref{eq:chiB}, that however without the
assumptions about the windows properties is at least tedious. Second, even if such an expression is found, this will lead to mixing kernels, which will not be numerically computable from the 'first principles', as in e.g., Eq.~\eref{eq:kernelScalar}, as they will involve the product 
of three functions : $P_{\pm2}$ multiplied by either $W_0$ or $M$, and by $W_{s=1,2}$, therefore leaving time consuming Monte Carlos as the only viable option for their estimation.

\subsection{\kn-approach}
\label{subsec:kn}
\subsubsection{Theoretical description}
Another way of estimating the $\chi^B$ field is by generalizing its definition to the cut-sky case. This can be done straightforwardly by modifying Eq.~\eref{eq:chiBdef} as follows,
\begin{equation}
\tilde{\chi}^B(\vec{n}) = \frac{i}{2}\left[\bar\partial\bar\partial M\,P_{2}(\vec{n})-\partial\partial M\,P_{-2}(\vec{n})\right],
\label{eq:chiBdefCut}
\end{equation}
where as usual $M$ stands for a binary mask and the tilde over the $\chi$ symbol is used to emphasize that at least in principle this is a different object than the true $\chi^B$ field defined on the cut sky, {\it i.e.}, $M\,\chi^B$. 
We note however that as long as $M$ is constant (and for simplicity assumed to be equal to $1$), {\it i.e.}, in the interior of the observed patch, the two fields are indeed identical $\tilde{\chi}^B(\vec{n})=\chi^B(\vec{n})$. In principle the only problem arises therefore at the patch edges. As proposed in Ref. \cite{kim_naselsky_2010} one could use this observation to reconstruct the true $\chi^B$ field everywhere
with an exception of the boundary layer. The problem becomes then technical and boils down to a question how to calculate the derivatives required by such a procedure. \cite{kim_naselsky_2010} propose to
do it in the harmonic domain and use semi-analytic formulae of \cite{zaldarriaga_2001} to represent the derivatives via convolutions of some geometrical kernels.
Given that the mask is abruptly falling from $1$ to $0$ at the edges, it is not surprising, that such a procedure leads to significant oscillatory behavior at the edges, which extends well within the center part of the
observed patch. This is a result of the necessity of imposing a finite band-limit on all harmonic decompositions performed as part of this procedure, even if the considered functions, with an abrupt jump does not
have such a limit. Such a band-limit is directly related to the pixelization used to represent the polarization fields.
This has two practical consequences. First, a robust criterion has to be found deciding which pixels are to be retained, i.e, which are sufficiently clean of any $E$-mode contamination, 
second, the loss of area is expected to be rather significant. We refer the reader to \cite{kim_naselsky_2010} for more details of this specific implementation.

A more robust approach would either invoke different ways of calculating the derivatives, {\it e.g.}, as proposed by \cite{bowyer_2011}, or introducing in Eq.~\eref{eq:chiBdefCut} a smooth apodization, $W$, in place of the binary mask, $M$.
This second option was proposed by \cite{kim_2010} and this is the one we implement in this work. The apodization could alleviate the pixelization effects described earlier by truncating the band limit
of the apodized polarization field, so the harmonic domain derivatives perform better.
Such a window would need to have a central region, where $W$ is constant (and equal to $1$) before smoothly rolling off at the edges. As in the case of the binary mask only in this central region the reconstructed
$\tilde{\chi}^B$ field would coincide with the true one and would be used for the power spectrum estimation.

 The main advantage of such a technique is that it provides a clear criterion which pixels to retain or to reject. Nevertheless, it does not solve completely the pixelization effects as pixels inside the central area can be affected by the pixel-induced leakage but this time originating from the contour around this central area.  However, and as numerical results shown in \cite{kim_2010} suggest, the pixelization effects at the inner contour are mitigated by the fact that $W$ is continuous as compared to the pixelization effects induced by considering the non-continuous binary mask. 

Hereafter, we will use this second approach and  apply a sky apodization to the polarization field. We will then use Eq.~\eref{eq:chiBdefCut} but with a mask, $M$, replaced by a window, $W$, to calculate
$\tilde{\chi}^B$ and later, the true $\chi^B = \l.\tilde{\chi}^B\r|_{M_{\chi^B}}$ where, $M_{\chi^B}$ is the binary mask built from the kept-in-the-analysis pixels, {\it i.e.}, pixels for which $W$ is essentially constant.

\subsubsection{Numerical implementation}
 The numerical implementation of this approach consists then in two main steps, which need to be first applied to simulated and later actual data. The Monte Carlo simulations are employed to
 select optimal windows for a given problem.
 
\medskip
\noindent$\bullet$~{\it Step 1}:
\medskip

 We calculate the apodized $\tilde{\chi}^B$ field for a selected window, $W$. This involves performing numerical derivatives of the available polarization fields, $P_{\pm 2}$ and
 those are performed in the harmonic domain. In this work we use a family of arch-sine windows as defined in~\cite{grain_etal_2009} with an apodization length which is to be tuned via Monte Carlo simulations.
 The criteria we use in the apodization length optimization process are the level of the $B$-spectrum bias and  variance.

\medskip
\noindent$\bullet$~{\it Step 2}: 
\medskip

We compute the $B$-mode power spectrum from the precomputed $\tilde{\chi}^B$ field. The spectrum is computed using only the trimmed, central part of the
available patch, $\l.\tilde{\chi}^B\r|_{M_{\chi^B}}$, which can be further  apodized, if needed, and follows the general pseudo-spectrum method framework.
Hereafter, following~\cite{kim_naselsky_2010} we will neglect possible leakages from the $E$-spectrum and use the scalar kernel  as also used in the \zb-approach, Eq.~\eref{eq:kernelScalar}.
We note however that unlike in the \zb-method the leakage in this approach can be more pervasive affecting even the most central areas of the patch and therefore never fully removed via simple area trimming.
 For this reason one may ponder whether a more appropriate
kernels can not be derived, which could account for these effects.
The answer, which we discuss in more detail in Appendix~\ref{app:knkernel}, is that such kernels would need to be evaluated numerically and be necessarily very costly. We will therefore
only consider the simplified case in this work.

\subsubsection{Sky apodization}
\label{sect:skyap}
The sky apodization and masking needs to be performed on three different stages in this approach.
First, we need to apodize the maps before computing the $\tilde{\chi}^B$ field. Then we need to mask pixels, which are expected to be contaminated by the residual $E$-to-$B$ leakage.
Finally, we may want to apodize the reduced $\tilde{\chi}^B$ maps to localize better bin-to-bin correlations of the recovered $B$ spectrum.

Unlike in the case of the \sz- and \zb-techniques, one cannot derive here some optimal windows from 'first principles'.  Instead for the sky apodization required for the computation of $\tilde{\chi}^B$ we use a family of the arch-sine analytic windows, proposed in~\cite{grain_etal_2009}, and resort to Monte Carlo simulations to optimize their apodization length. 
In this optimization procedure we always trim all the pixels within the boundary layer of $W$, i.e., where it is not constant, as these are the pixels, which are unavoidably affected by the 
$E$-to-$B$ leakage, and we use only the remaining ones for the spectrum estimation. Clearly, there will be still some level of the $E$-mode power in the map left over after such a trimming procedure, mostly due to pixel induced $E$-to-$B$ leakage. 
The level of this leakage depends on the assumed apodization length, becoming slower for its larger values, and the MC simulations are then used to find the smallest value
of the latter ensuring a sufficiently low level of the leakage. This will at the same time maximize the sky
area, given the acceptable leakage requirement, left for the final spectrum determination and therefore ensure that the spectrum variance is the smallest.

\subsection{Brief appraisal}
\label{sec:comparison}
The three methods considered in this work  can be introduced within a common framework based on the $\chi^B$ field concept as has been done in this Section and demonstrated to be all rather closely related. The fact, which may be  potentially somewhat surprising given their original derivations. 

The two first methods, \sz\ and \zb, in the renditions as considered in this paper are clearly equivalent on 
the analytical level, if the apodizations employed in both these cases are related to each other as, $W_{ZB} = \sqrt{W_{SZ}}$, and $W_{SZ}$ fulfills strictly the boundary conditions. The differences between these two approaches are therefore only in their numerical implementations and approximations which they imply. Both these methods suffer due to pixelization issues, in particular arising due to a need to compute numerical
derivatives, and which give rise to a residual contamination of the $B$-spectrum with the $E$-mode power.
The \sz-method requires only derivatives of the window functions,  therefore, at least in the cases when these are given analytically, it is possible to estimate the non-diagonal coupling kernel, $K^{(-)}$,
and correct for some of those effects. Such corrections are more difficult in the case of the \zb-approach, where the non-diagonal kernel would have to be estimated completely numerically.
The \sz-method can potentially offer more freedom for an optimization of the $B$-spectrum variance as estimated for realistic noisy maps as the boundary conditions on the applied apodizations can be relaxed 
leading to an increase of the signal variance related to allowing for some $E$-to-$B$ leakage but a decrease of the total, signal+noise, one. At the same the off-diagonal, polarization mode coupling kernels
can be readily calculated and the estimated $B$-spectrum unbiased.

The \kn-approach can be looked at as an approximation of the \zb-method. Indeed the first term on the rhs of Eq.~\eref{eq:chiBdef} used by the \zb-method coincides with the rhs of Eq.~\eref{eq:chiBdefCut} (replacing $M$ by a sky apodization $W$), which defines the first step of
the \kn-approach. We refer to App. \ref{app:knzb} for a detailed discussion. The  contributions of the extra three terms in Eq.~\eref{eq:chiBdef} are localised around the patch boundary and removed in the \kn-method by trimming the boundary layer, which is retained and used for the power
spectrum estimation in the case of the former method.  For this reason we may expect
that the performance of the \kn-method should be inferior to both the \zb- and \sz-approaches, which in turn we could expect to be nearly equivalent.
In turn, the \kn-method may appear as the most straightforward on the implementation level and therefore attractive at least at first stages of the analysis.

\section{Numerical experiments}
\label{sect:results}

\subsection{Experimental set-ups}
\label{sec:exp}
For numerical investigations, we define two fiducial experimental setups. Though idealized, they are chosen to reflect the general characteristics of forthcoming CMB experiments dedicated to $B$-modes detection. Those characteristics which crucially impact on the angular power spectrum reconstruction are the noise level, the beam width and a peculiar sky coverage.

\begin{figure}
\begin{center}
	\includegraphics[scale=0.3]{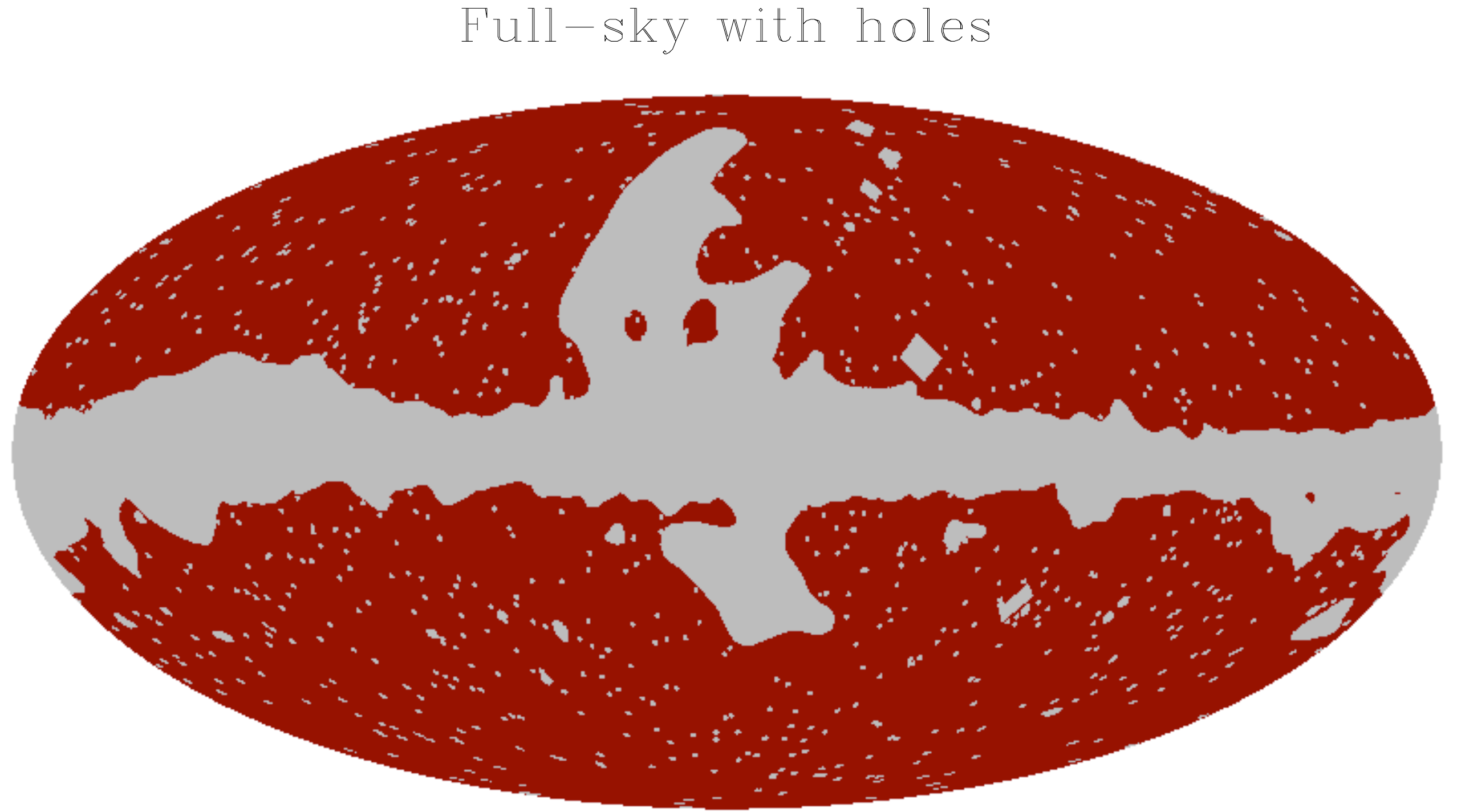} \\
	\vspace*{0.25cm}

	\includegraphics[scale=0.3]{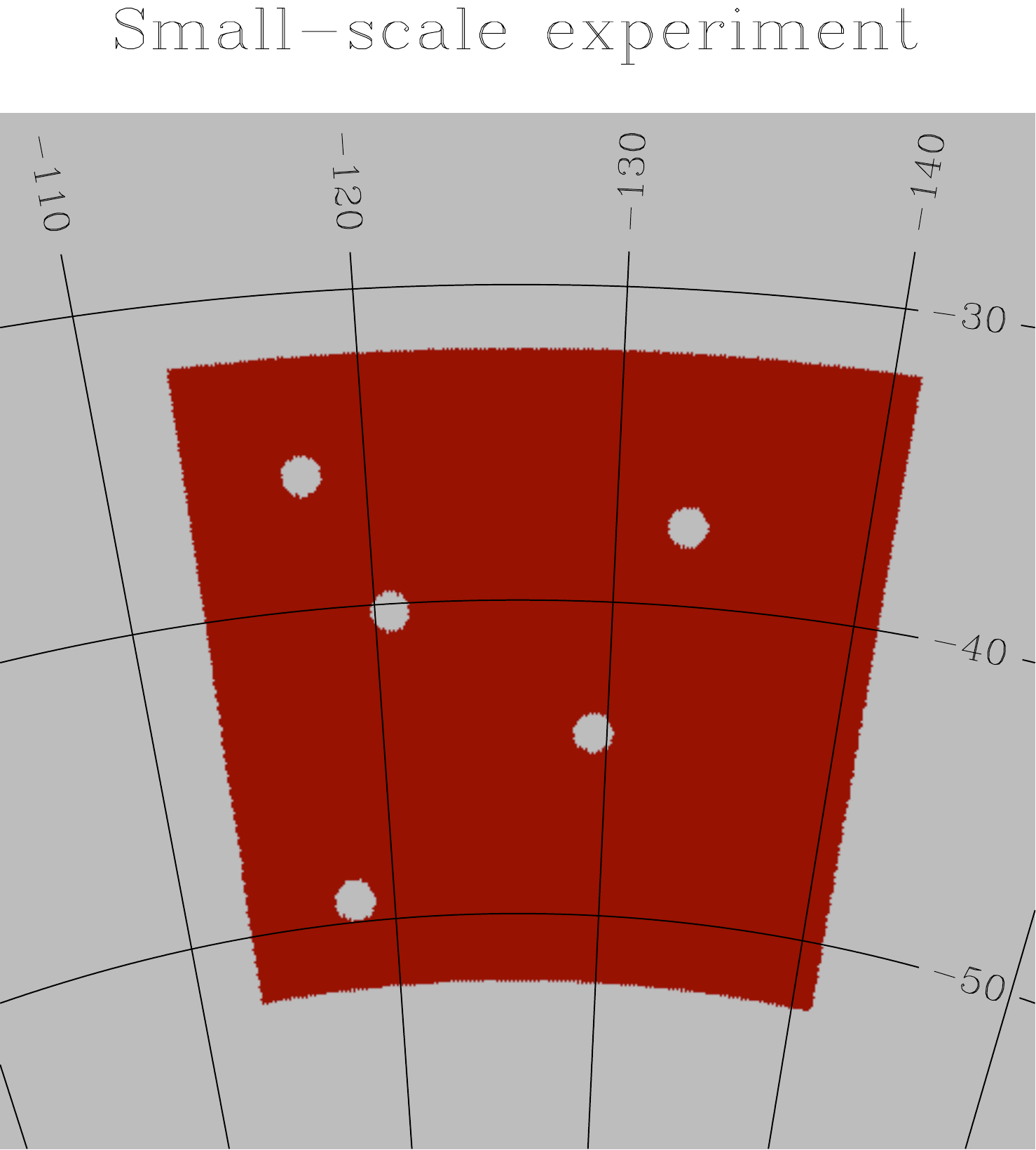}
	\caption{Sky areas as observed by the fiducial satellite-like experiment (upper panel) and for the balloon-borne, small-scale experiment (lower panel) as considered in this work. The sky coverages are respectively $\sim71\%$ and $\sim1\%$ of the total celestial sphere. For the satellite experiment, the mask is a combination of the galactic mask $R9$ and the point-sources catalog used for polarized data in \textsc{wmap} 7yr release. Only the latter mask is used for the balloon-borne case.}
	\label{fig:masks}
\end{center}
\end{figure}

We first consider the case of a possible satellite experiment aimed at $B$-mode detection. For such an experiment, we relied on the {\sc{epic}}$-2m$~\cite{bock_etal_2008} 
specifications for the noise level and the beam width, setting these to $2.2~\mu K$-arcmin for the noise level and $8~$arcmin for the beam width. For the peculiar sky coverage of such a 'nearly full-sky' experiment, we consider the galactic mask $R9$ used for polarized data in \textsc{wmap} 7yrs release (see \cite{wmap_7yr}) adding the point-sources catalog mask. So we obtain a $\sim71\%$ sky coverage patch showed in the lower panel of Fig.~\ref{fig:masks}. Throughout this work we use Healpix pixelization scheme~\cite{gorski_etal_2005}. Here the pixel size is $\sim7$ arc minutes, {\it i.e.} $N_\mathrm{side}=512$.

Second, we consider the case of balloon-borne experiment inspired by the ongoing {\sc ebex} experiment \cite{britt_etal_2010}. The noise level and the beam width are respectively set equal to $5.75~\mu K$-arcmin and $8~$arcmin. The observed part of the sky covers $\sim1\%$ of the total celestial sphere and its peculiar shape is displayed on the upper panel of Fig.~\ref{fig:masks}. It consists of a square patch of an area  of $\sim400$ square degrees including holes to mimic polarized point-sources removal. In such a case, we choose $N_\mathrm{side}=1024$ corresponding to a pixel size of $\sim3.5$ arc minutes.

\subsection{Simulations}
\label{sec:sim}
We numerically implement the three techniques described in the previous section and test their respective efficiency with Monte-Carlo simulations.
We investigated the full performances of those approaches from the perspective of $B$-mode power spectrum reconstruction and therefore incorporate noise with the level as stated in Sec. \ref{sec:exp}. To simulate the CMB sky, the input $E$-mode signal is that of the cosmological model with parameters as given by the WMAP 7yrs data \cite{larson_etal_2011} and the input $B$-mode includes lensing and primordial $B$-modes with $r=0.05$ (Our convention for $r$ follows the WMAP convention: $r=\mathcal{P}_\mathrm{T}(k_0)/\mathcal{P}_\mathrm{S}(k_0)$ with $\mathcal{P}_\mathrm{S(T)}$, the {\it primordial} scalar(tensor) power spectrum and $k_0=0.002$~Mpc$^{-1}$ the pivot scale).

We will  assume that two identical maps are always available with the same level of the homogeneous noise in each of them, which is taken to be  uncorrelated between the two maps and use their cross-spectra and their variance to compare different approaches. We calculate the latter
with help of  Monte Carlo simulations and use as a common reference an estimation of the variance based on simple mode-counting and given by
\begin{equation}
	\mathbf{\Sigma}_{\ell\ell'}=\frac{\delta_{\ell,\ell'}}{(2\ell+1)f_\mathrm{sky}}\left[\left(C^{B}_\ell\right)^2+\left(C^B_\ell+\frac{4\pi}{N_\mathrm{pix}}\frac{\sigma^2_p}{B^{2}_\ell}\right)^2\right],
\end{equation}
where $B_\ell$ is the beam function and $\sigma_p$ -- the noise per pixel. This formula applies to a {\it cross-spectrum} between two maps and assumes that the noise of the two maps is uncorrelated and its level per pixel is given by $\sigma_p$. This na\"\i{v}e mode-counting is bound to underestimate the variance in our study cases and is therefore used only as a lower limit. 

An  effective, observed fraction of the sky, $f_\mathrm{sky}$, depends on an assumed apodization  and therefore will be in general different for each of the methods considered here and may vary from a bin to a bin. 
For definiteness hereafter as a reference we will use its value computed assuming only binary mask, $M$. Such a choice, in terms of the Fisher errors leads to the lowest variances.


\section{Results: satellite case}

\subsection{Standard pseudo-spectrum method}

The major advantage of the satellite experiments is their ability to measure the sky signals on the largest angular scales, and therefore having potential
to constrain their power spectra all the way to the lowest multipoles.
Indeed, the simple Fisher variance formula introduced earlier seems to suggest that this should be possible if only the sky coverage is sufficiently large. Though this formula neglects the leakage it seems only natural to expect that it 
should be small for nearly full sky maps, and therefore should lead to subdominant effects as compared to other uncertainties, e.g., cosmic variance.
 
In this section we confront these expectations against realistic simulations within the paradigm of pseudo-spectrum methods. In this context, if the leakage is indeed small, we may expect that even the standard pseudo-spectrum 
technique could perform sufficiently well assuring precision comparable to that of the other methods, which explicitly invoke some leakage correction, and not that far off the Fisher predictions.
Below we therefore start from a discussion of  the standard pseudo-spectrum technique.

\subsubsection{Leakage}
We quantify the level of the $E$-to-$B$ leakage using standard pseudo-spectra calculated in the case of simulations with no input $B$-mode power and, which would have been zero had there been no leakage at all. These are denoted hereafter as $\tilde{C}^{E\rightarrow B}_\ell$. We compare these pseudo-spectra with those calculated assuming no input $E$-mode power, denoted $\tilde{C}^{B\rightarrow B}_\ell$, and therefore expressing the pseudo-power of the genuine $B$-modes.
These pseudo-spectra are shown in Fig.~\ref{fig:pseudostd},  which displays  $\tilde{C}^{E\rightarrow B}_\ell$, upper curve, and $\tilde{C}^{B\rightarrow B}_\ell$, lower curves, computed for three different values of $r=0.1,~0.05, ~0.01$.  
 Clearly,  the leaked power, $\tilde{C}^{E\rightarrow B}_\ell$, dominates over the true $B$-modes at least up to $\ell \sim 700$. We therefore conclude that the leakage is by far not insignificant even in the satellite case.
 
Furthermore, if we take a ratio of  $\tilde{C}^{E\rightarrow B}_\ell$ and $\tilde{C}^{B\rightarrow B}_\ell$ as a measure of the magnitude of the leakage we find that its values are within a factor of $2$ from those obtained for the small-scale experiment considered later on, indicating that the leakage amount in both cases is in fact comparable, even if the latter experiment covers roughly $\sim 71$ smaller sky area than the former. 
 
This  demonstrates that it is not merely sky area which matters as far as the leakage is concerned. In fact, the gain in the sky area in the case of the satellite experiment considered here comes at the price of a significantly more complex and longer perimeter, effects of which, \citep[e.g.][]{bunn_etal_2003}  offset the sky area advantage. We note that though we may attempt to simplify the boundary of the Galactic mask to suppress
the leakage, this is more difficult to be done with the point sources, which indeed seem to provide the major contribution to the observed
level of the leakage.

\subsubsection{Variance}

The large leakage found present on the pseudo-spectrum level will inevitably lead to excess variance of the $B$-mode spectrum estimate. These are 
depicted in Fig.~\ref{fig:varstd}, where variances computed assuming three different apodizations are shown. We see that in either case no meaningful
constraints on the lowest multipoles, $\ell \simlt 30$, can be set at least as long as no binning is applied. These results demonstrate that for realistic observations the standard pseudo-spectrum method can not ensure sufficient precision for the largest angular scales
and some alternatives, explicitly correcting for the leakage, need to be considered instead, as we do so in the next section.

\begin{figure}
\begin{center}
	\includegraphics[scale=0.525]{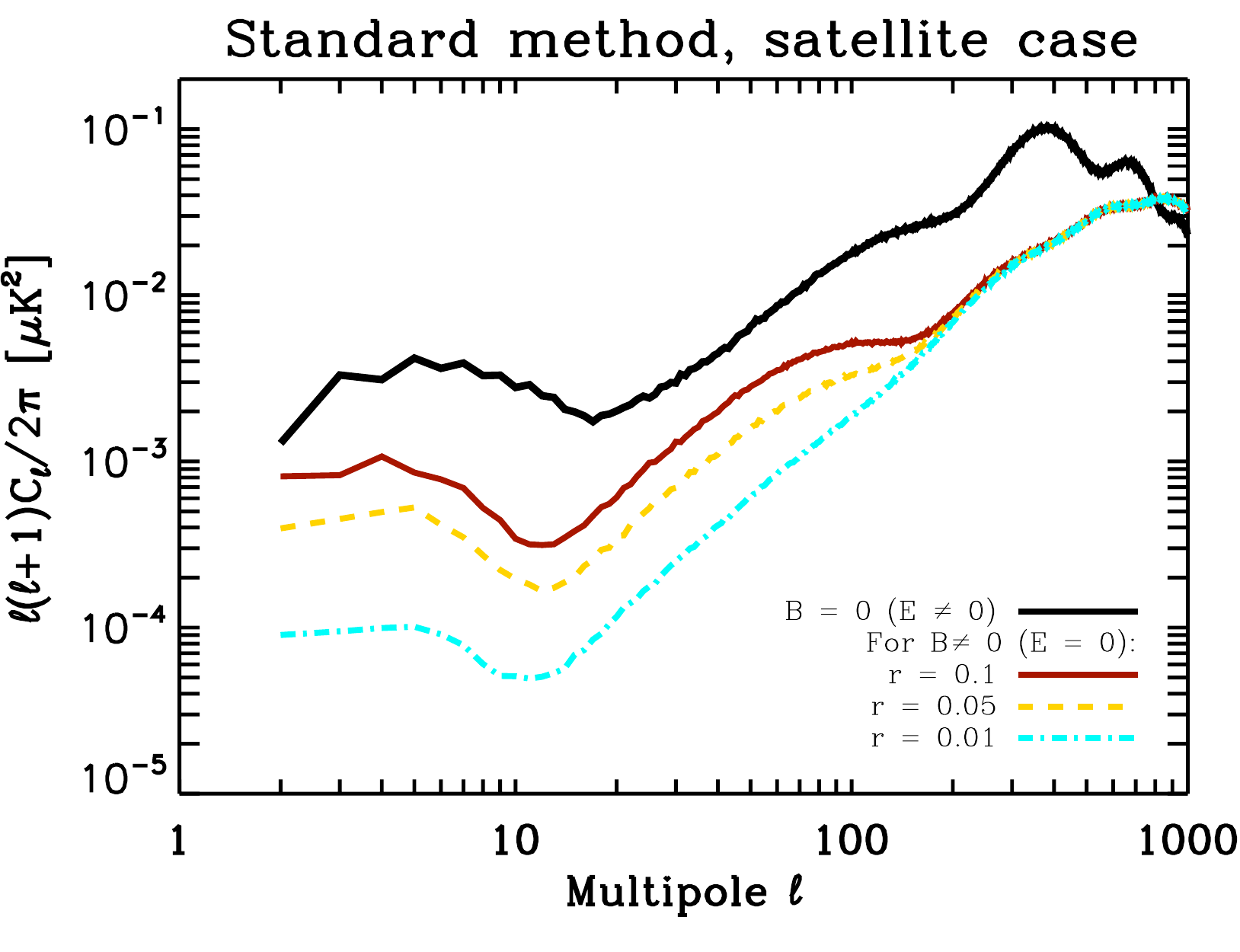}
	\caption{Contribution of $E$-modes (black curve) and $B$-modes (colored curves) to the $B$-modes pseudo-$C_\ell$ for the case of a satellite mission. This measures the relative amount of $E$-mode leaking into $B$ if one does not correct for such leakages. The corresponding mask is depicted in the upper panel of figure \ref{fig:masks}.}
	\label{fig:pseudostd}
\end{center}
\end{figure}

\begin{figure}
\begin{center}
	\includegraphics[scale=0.525]{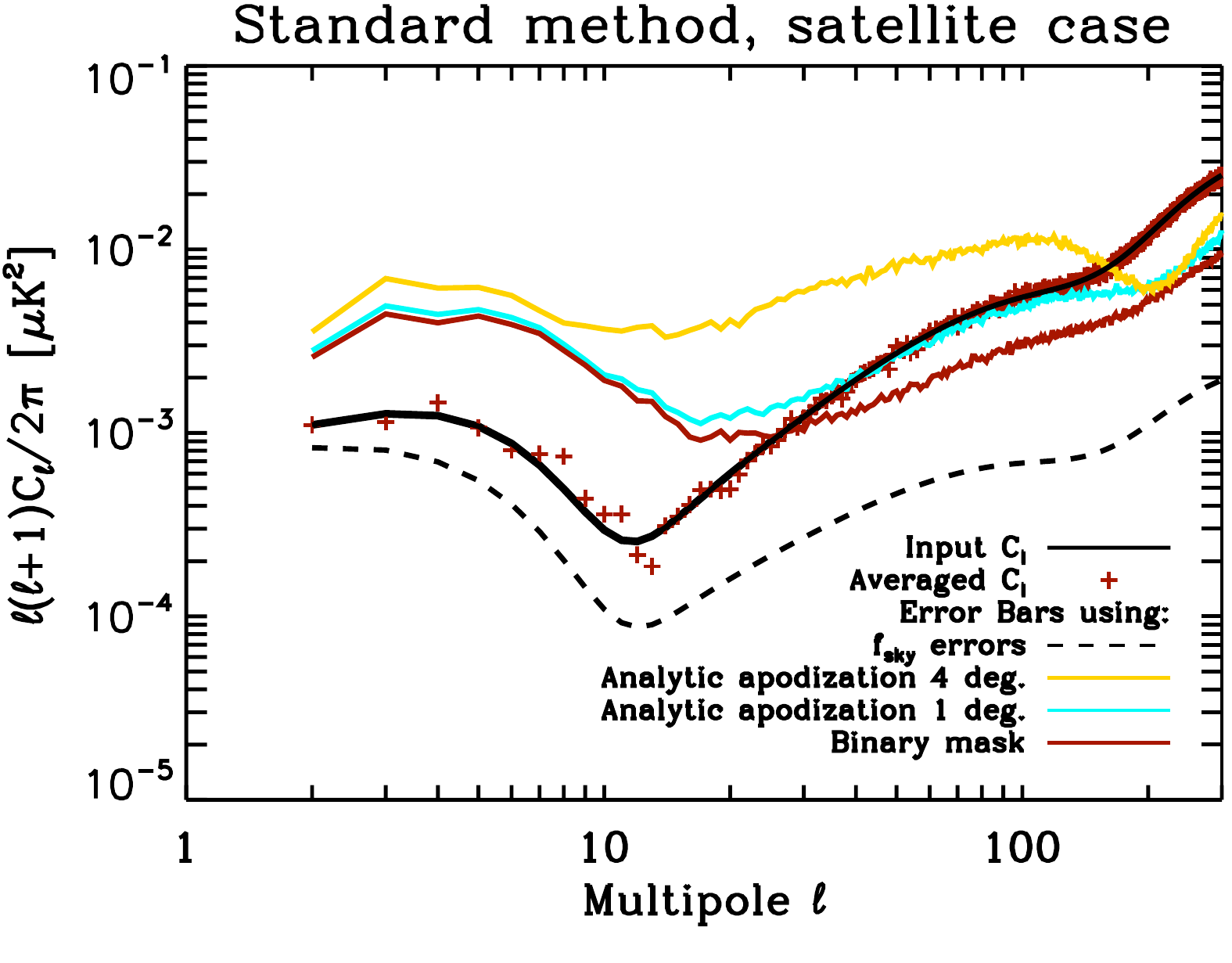}
	\caption{Reconstruction of the $B$-modes power spectrum for $r=0.05$ in the case of a satellite mission ($f_\mathrm{sky}=71\%$) using the {\it standard} pseudo-spectrum approach. The noise level is $2.2\mu K$-arc minute and the beam is 8 arc minutes. The solid-black curve is the input power spectrum and the dashed-black curve is the Fisher estimate of the error bars. The red crosses stand for the recovered power spectrum and the colored curves stand for the error bars from MC simulations using different apodization length for the sky apodization applied to the polarization maps.}
	\label{fig:varstd}
\end{center}
\end{figure}

Fig.~\ref{fig:varstd} also shows a $B$-mode spectrum averaged over all performed MC simulations. It is unbiased, as expected, given that we include
explicitly in the calculations the off-diagonal coupling kernel, $K^{(-)}_{\ell\ell'}$, correcting the spectra on average for the $E$-mode power leaked 
to $B$. In practice, we find however that a special care needs to be taken 
while calculating this kernel to ensure the absence of the bias.
This is because the leaked power is indeed grossly dominant over that genuine B-mode, see Fig.~\ref{fig:pseudostd}, setting very demanding constraints on the precision of the kernel. For instance, the good agreement shown in Fig.~\ref{fig:varstd} has been only obtained, when
we minimized the spurious contributions due to the pixelization coming specifically from the polar caps by rotating the sky map so those have been hidden in the regions excluded by the employed mask.
The residual scatter at its low-$\ell$ end is just a result of the insufficient number of simulations and the huge variance displayed by the standard pseudo-spectrum estimator on these scales.

The good overall agreement of the averaged spectrum with  the theoretical spectrum used for the simulations validates our MC-based predictions for
the variances.
 

\label{sssec:pcg}
\begin{figure}
\begin{center}
	\includegraphics[scale=0.525]{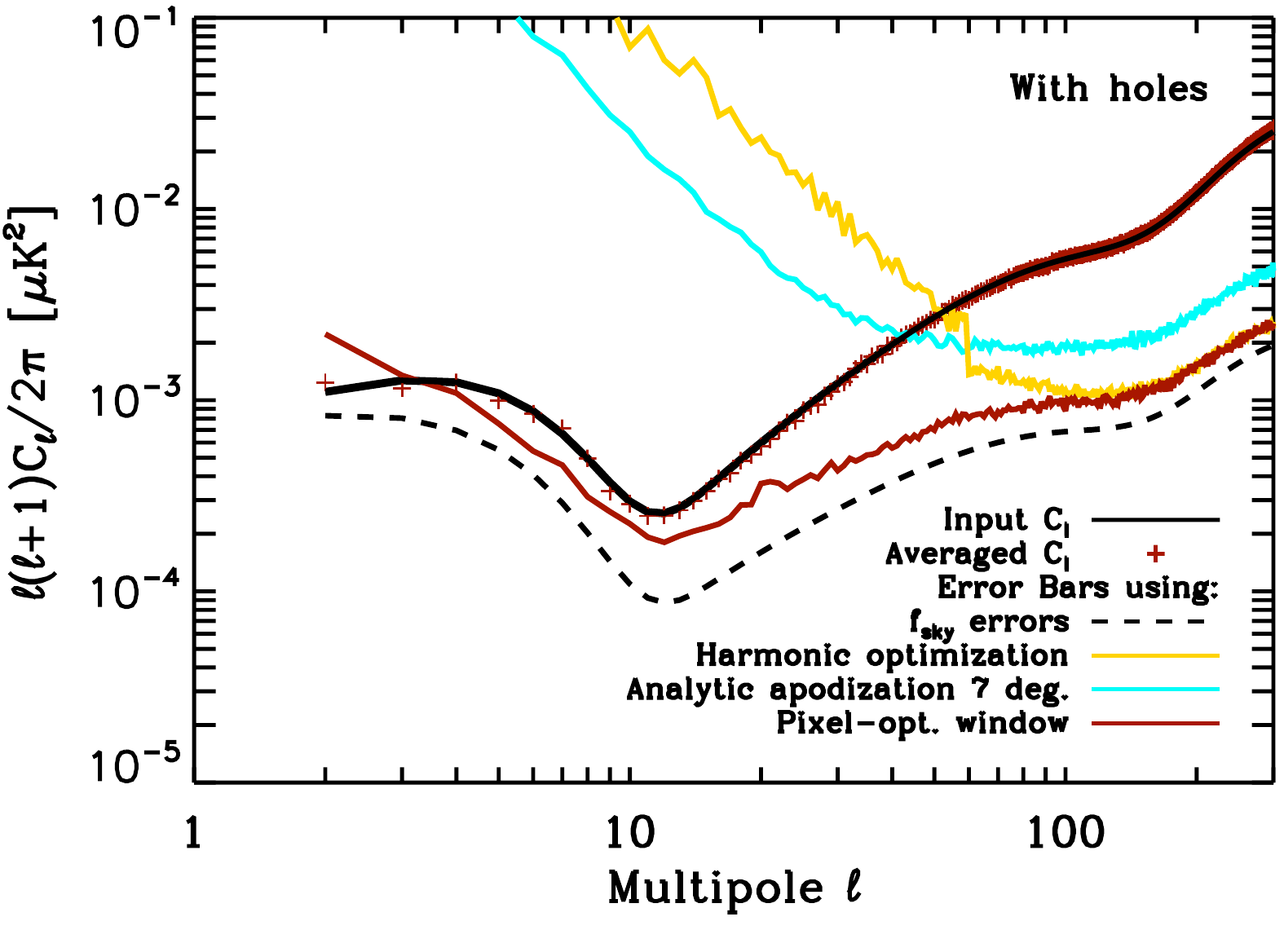}
	\includegraphics[scale=0.525]{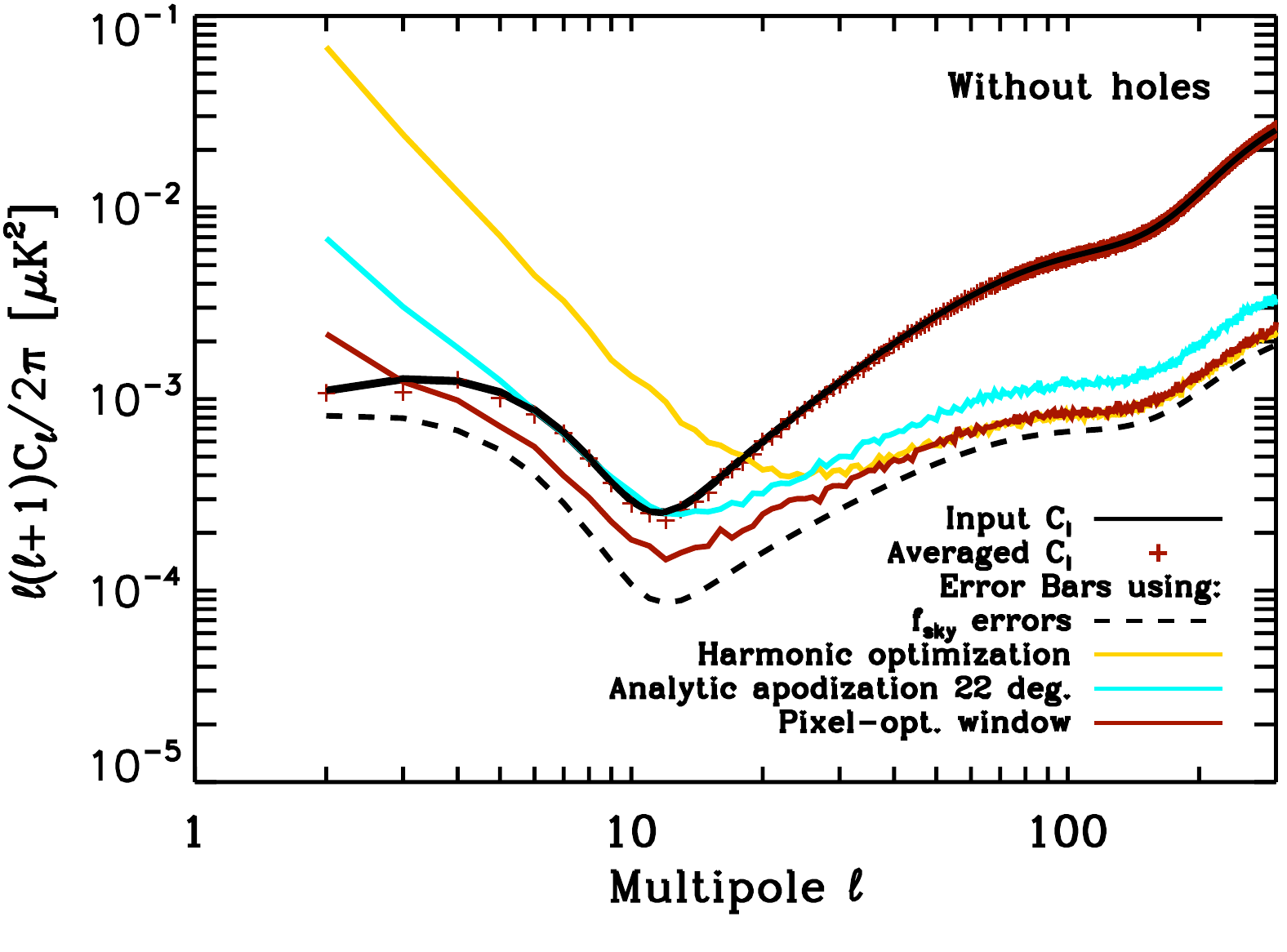}
	\caption{{\it Upper panel}: Reconstruction of the $B$-modes power spectrum for $r=0.05$ in the case of a satellite mission ($f_\mathrm{sky}=71\%$) using the \sz- pseudo-$C_\ell$ approach and using three types of sky apodizations. The noise level is $2.2\mu K$-arc minute and the beam is $8$ arc minutes. The solid-black curve is the input power spectrum and the dashed-black curve is the Fisher estimate of the error bars. Red crosses are the recovered power spectrum. The solid-yellow curve stands for error bars on $C^B_\ell$ recovery using a variance-optimized apodization forcing the boundary conditions and the relationship derivative to be fulfilled (computation in the harmonic domain). The solid-red curve corresponds to the error bars using a variance-optimized apodization relaxing those conditions (computation in the pixel domain.) The solid-cyan curve shows the error bars as obtained using an analytic sky apodization with $\theta_\mathrm{apo}=7$ degrees. {\it Lower panel}: Same as upper panel but considering the galactic mask only and not the holes. The sky coverage is $73$\%. The apodization length for the analytic sky apodization is $22$ degrees.}
	\label{fig:varsz}
\end{center}
\end{figure}

\subsection{Leakage-correcting methods}

\subsubsection{Apodization}
\label{sec:apo}

The results described above demonstrate that the standard approach is not suitable for the low-$\ell$ recovery of the $B$-mode spectrum even for
the nearly full sky experiments. Therefore, if such a goal is achievable at all with a pseudo-spectrum method, it would have to be a method, which tackles the leakage problem case-by-case, as do the three methods discussed earlier. It is important however to emphasize that the suppression of the $E$-to-$B$ leakage
in these methods comes at a price as the corrections they invoke may affect the variance of the recovered spectrum. Consequently, this variance will 
not be in general close to the variance of the $B$-mode spectrum as obtainable in the standard pseudo-spectrum approach in a case, when the CMB $E$-mode power, and therefore the leakage, is set artificially to zero, as one could ideally hope for. Instead there will be typically an extra contribution to the variance, not due to the leakage anymore, as it is explicitly treated for, but from removal of  part of the information as resulting from the leakage correction procedure. 

This in principle calls for some optimization procedure between the level of the leakage
and the bias (at least for some of the methods studied here) and the variance of the recovered $B$-mode power spectrum. As the loss of the information is
related to the apodization and/or masking applied in these methods, and used sometimes on multiple stages,
such an optimization could be in general rather cumbersome to formalize and to date has been implemented in a systematic way only in the case of the \sz-approach~\citep[]{smith_zaldarriaga_2007}. In this method the estimated power spectrum is always unbiased and the variance level is uniquely determined by one -- if the boundary conditions and relations between different spin windows are strictly enforced -- or three window functions -- if the boundary conditions are relaxed and no relations between windows is imposed. In the latter case, one admits some level of leakage but tries to 
capitalize on the additional
freedom to gain on the resulting variance. In the past literature~\citep[e.g.][]{smith_2006, smith_zaldarriaga_2007, grain_etal_2009} a number of either ad hoc or optimized windows have been considered and shown to perform comparably at least in the simplest circumstances. In Fig.~\ref{fig:varsz} we show the variances
obtained with the \sz-method assuming a selection of windows in the case of our satellite set-up assuming presence of the masked point sources, 
upper panel, or not, lower.
We observe that there is huge disparity in the performance of the different windows in particular at the low-$\ell$ end of the spectrum.  The windows, which tend to impose the boundary condition, i.e., harmonic and analytic ones, perform
 significantly worse than the window for which these are relaxed,
i.e., the pixel-domain optimized window. Moreover, the variances in the former cases are often  significantly worse than those obtained in the case of the standard approach in particular at the low-$\ell$ end.

We can therefore conclude that not only the pixel domain optimized windows provide the best performance, at least out of the cases we have looked at here, but also that they are unique in ensuring essentially the same performance in the cases 
of the both masks considered here. For this reason we will use these windows, whenever applying the \sz-approach in the following.

We note that the pixel-domain computation of the optimized windows does involve significant computational resources, which are needed to solve iteratively large linear systems~\citep[][]{smith_zaldarriaga_2007}, for a number of $\ell$-bin, and which dominate the overall computational cost of
the approach.

The situation is more complicated in the cases of the other two methods as equivalent optimization procedures have not been proposed in their context. This is in part due to technical problems related to the dimension of the parameter space, which would have to be considered. We therefore do not attempt to 
devise such procedures in this work. Instead, in these cases we will apply simple analytic apodizations and demonstrate the dependence of the obtained results on their parameters. As these apodizations may not be optimal, it may be in principle possible to improve  on the results we derive in the following. However, we find that in general the results for these two methods are less sensitive to the apodization choices
than those derived in the case of the \sz-approach and therefore we do not expect the improvement to be significant and affect our conclusions.

We note that even with the proper optimization the determination of the low-$\ell$ multipoles, multipole-by-multipole is burdened with a significant
error. Indeed, the variance is comparable to the signal amplitude for $\ell \simlt 20$ and even larger than the latter for $\ell \simlt 3-4$. For this reason, in the following we will always bin the spectra even in the nearly full-sky case considered here. The choice of binning will be marked at the bottom of each plot as grey shaded boxes. The lowest bin will then span $\ell$ values from $2$ up to $20$.

\begin{figure}
\begin{center}
	\includegraphics[scale=0.525]{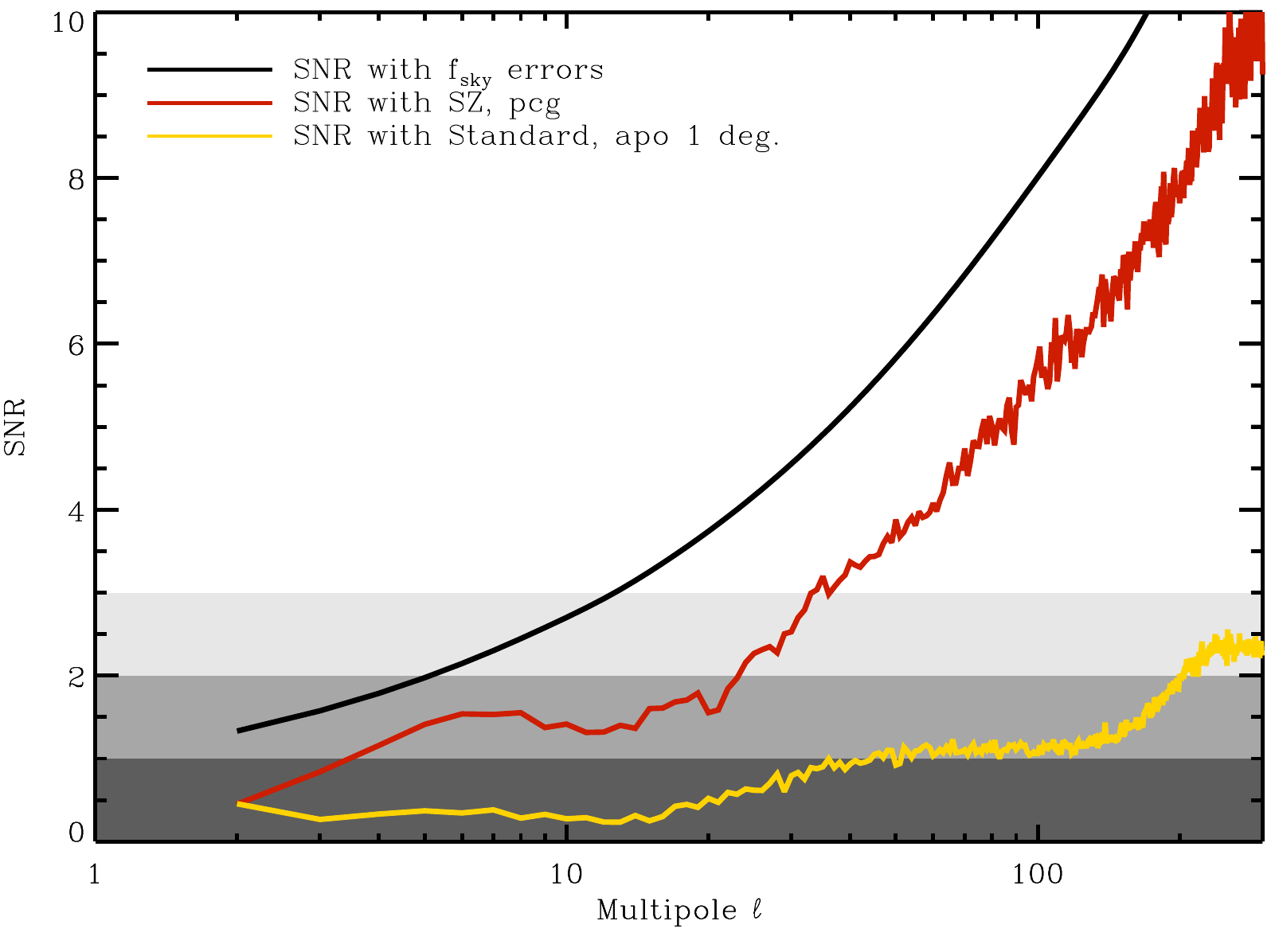}
	\caption{Signal-to-noise ratio $C^B_\ell/\sqrt{\mathbf{\Sigma}_{\ell\ell}}$. The red and yellow curves respectively stand for the \sz- and the standard pseudo-$C_\ell$ estimation. The black curve shows the SNR as obtained using the Fisher estimate of the uncertainties on the reconstructed $C^B_\ell$. The shaded grey areas highlights the 1-$\sigma$, 2-$\sigma$ and 3-$\sigma$ detections.}
	\label{fig:snr}
\end{center}
\end{figure}

The gain in using the \sz-approach as compared to the standard approach which does not correct for $E$-to-$B$ leakage is visiualized in Fig.~\ref{fig:snr}. It depicts the signal-to-noise ratio (SNR) of the $B$-mode angular power spectrum reconstruction, $C^B_\ell/\sqrt{\mathbf{\Sigma}_{\ell\ell}}$. The red curve stands for the SNR as obtained using the \sz-method while the yellow curve stands for the SNR as obtained using the standard pseudo-$C_\ell$ method. The black curve corresponds to an idealized SNR based on the Fisher estimate of the uncertainties. The shaded grey areas highlights the 1-$\sigma$, 2-$\sigma$ and 3-$\sigma$ detections. It is clear from such a figure that detecting the primordial component of $B$-modes, peaking at $\ell<100$, for a satellite-like survey requires to correct for $E$-to-$B$ leakage.

\subsubsection{Power spectrum recovery: bias and uncertainties}
\begin{figure}{ht}
\center
	\includegraphics[scale=0.5]{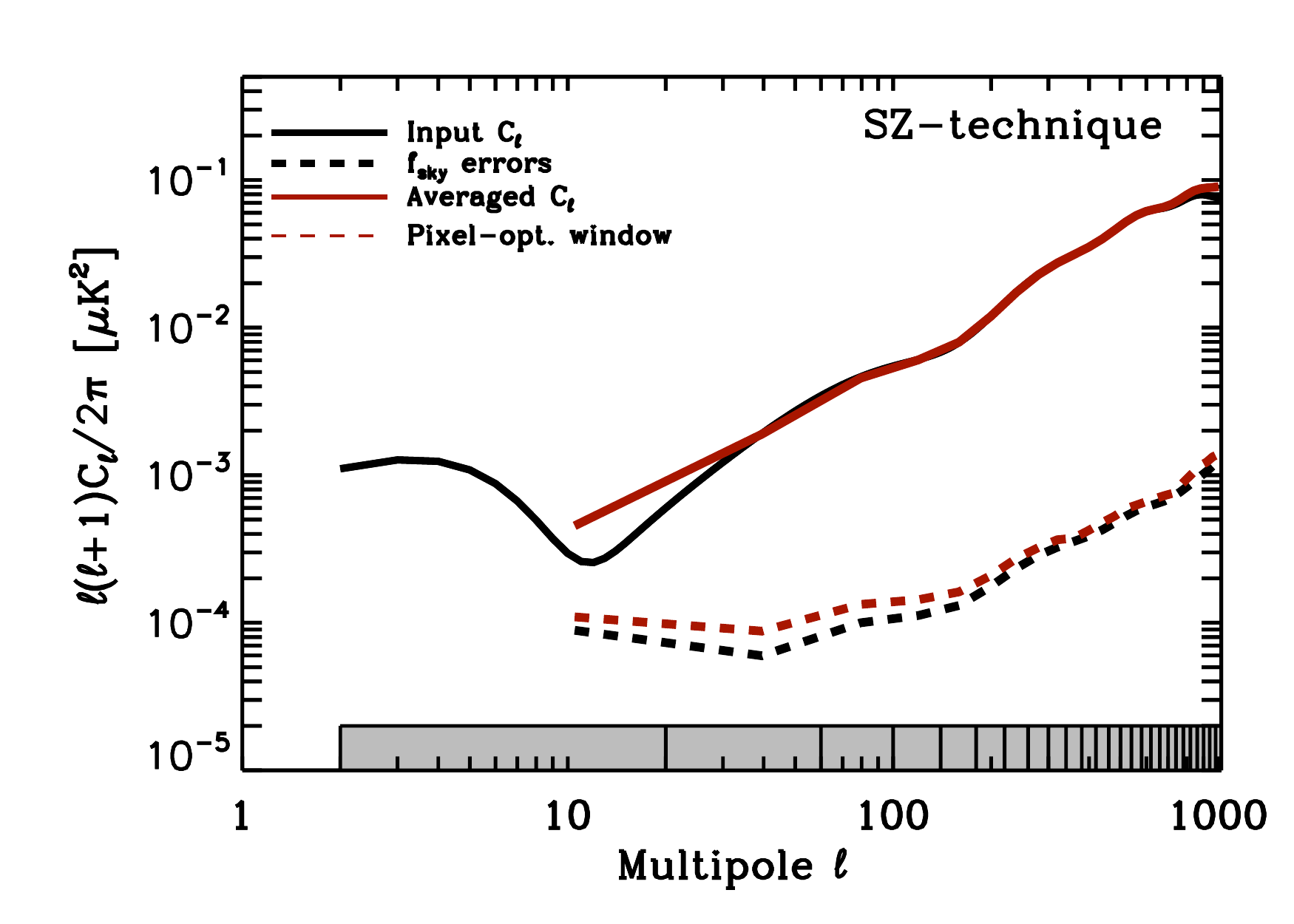} \\
	\includegraphics[scale=0.5]{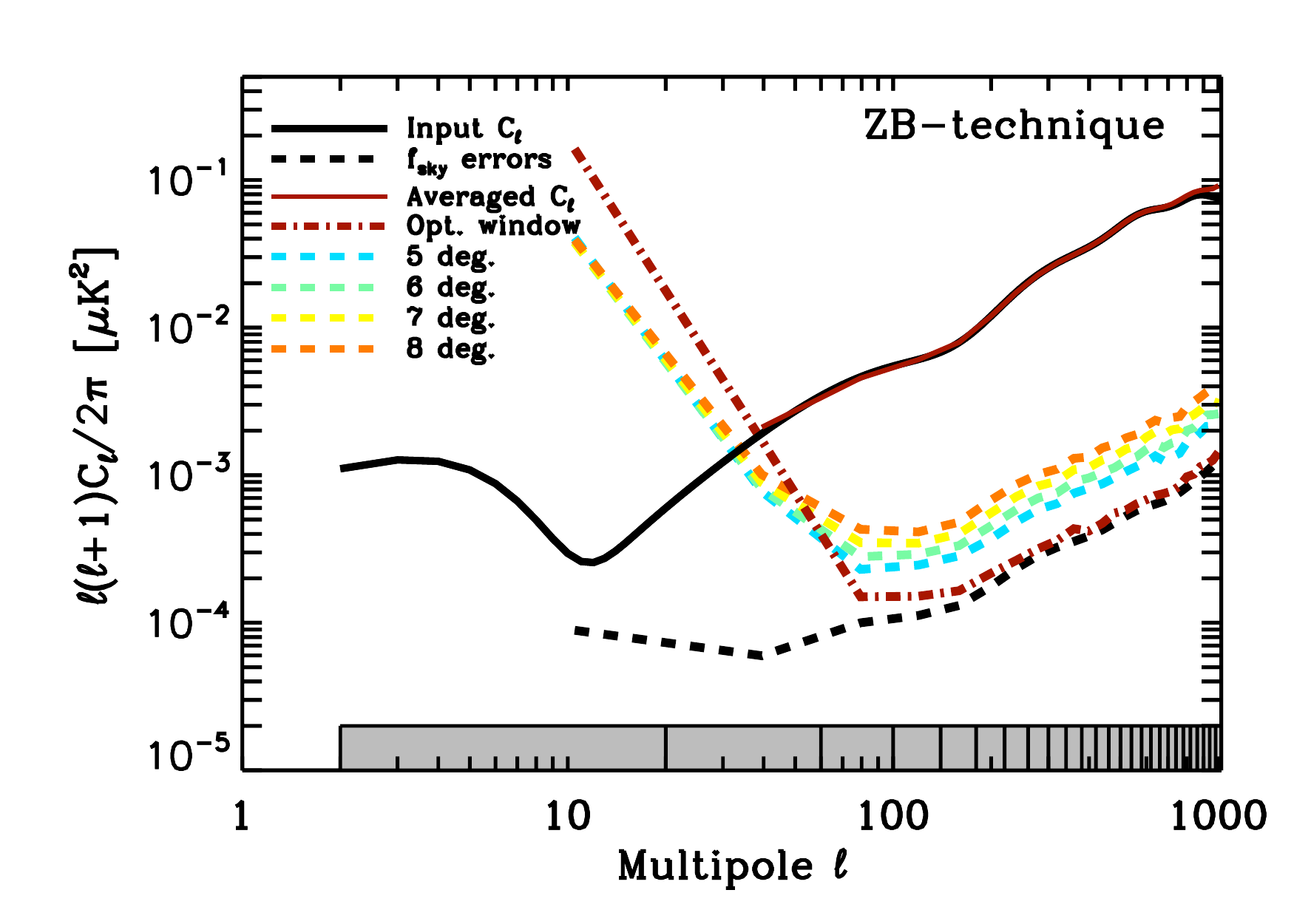} \\
	\includegraphics[scale=0.5]{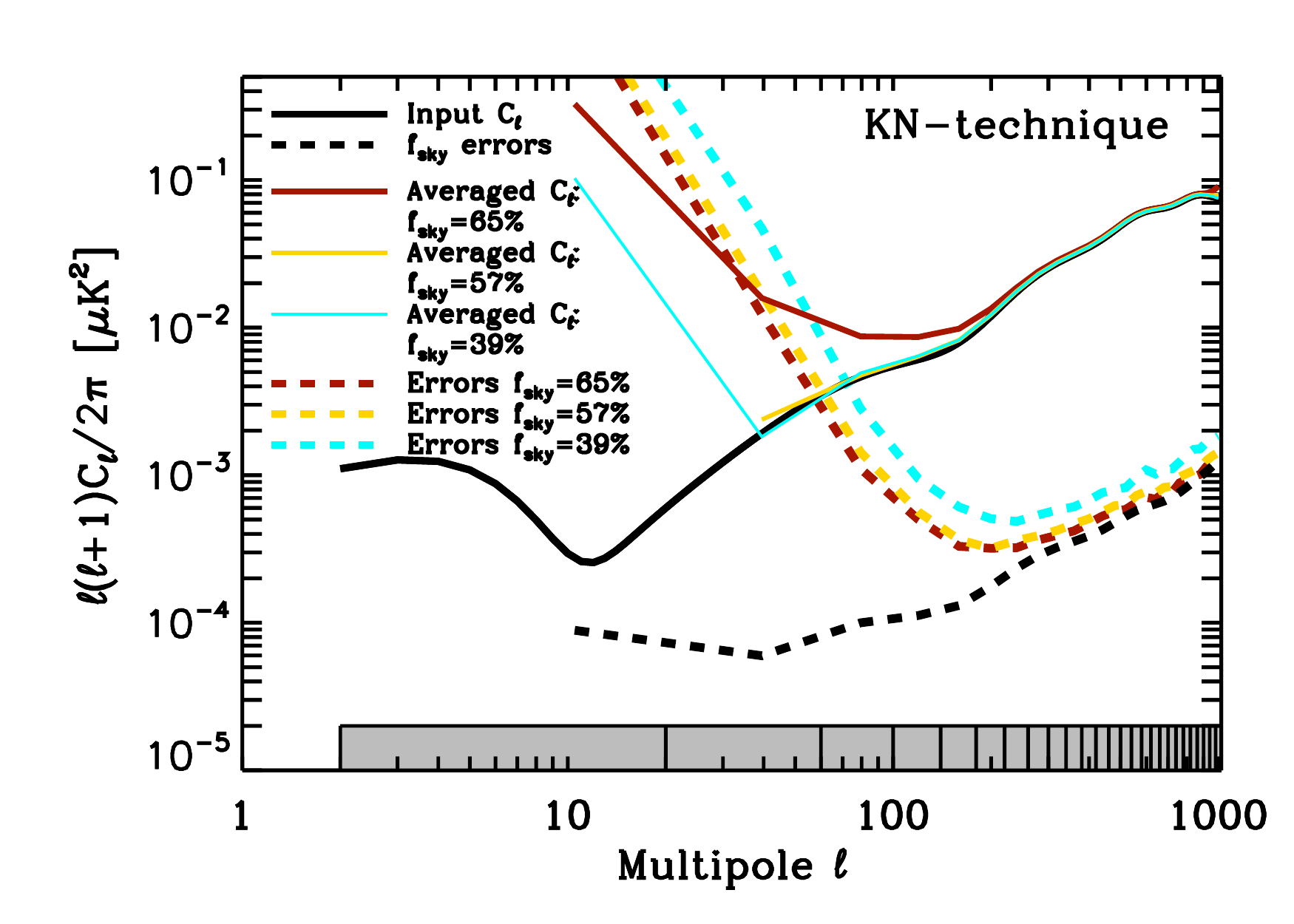}
	\caption{Power spectrum uncertainties on $B$-modes using cross-spectrum estimation for the case of a satellite-like experiment with holes mimicking point-sources removal ($f_\mathrm{sky}\sim71\%$). Upper, middle and lower panels are respectively for the \sz-, \zb- and \kn-methods. Grey shaded boxes represent the binning of the power spectra. The sky apodizations used for each techniques are described in Sec.~\ref{sect:formalism}.}
	\label{fig:cellwmap}
\end{figure}

The reconstructed $B$-modes angular power spectra and their uncertainties for each of the three above-described methods, are shown in Fig.\ref{fig:cellwmap}. The upper, middle and lower panels respectively stands for the \sz-, \zb- and \kn-techniques.  As explained in Sec.~\ref{sec:apo}, the angular power spectra are estimated for $\ell \in [2,1020]$ within multipoles' bands with bandwidth of $\Delta\ell = 40$. For each methods, we optimize the sky apodization to obtain the lowest error bars. 

The plotted solid-black curve stands for the input $B$-modes angular power spectrum while the solid-red is the estimated one, averaged over 500 simulations, which is build to be unbiased (we will discuss the results in practice for each method). The dashed black curve on each panels represents the mode-counting estimate of $C_{\ell}^B$'s uncertainties which are calculated as explained in Sec.~\ref{sec:sim}.
The dashed colored curves are the MC estimated uncertainties. Those estimated binned power spectra and their associated error bars are plotted at the central value of each bandpower. The width of the here-adopted bandpowers are depicted by the grey shaded rectangles. 

As already mentioned, the three pseudo-$C_\ell$ techniques {\it are theoretically built to} provide {\it unbiased} estimations of $C^B_\ell$. Nonetheless, due to numerical effects as the pixelization, the reconstructed $B$-modes may be biased. The bias and the uncertainties behaviors for each techniques are analyzed and compared hereafter.

(i)~{\it\sz-technique:}
As expected, our estimation of the $B$-mode angular spectrum is unbiased.
The window functions are optimized in the pixel domain leading to uncertainties very close to the mode-counting estimation throughout the entire range of angular scales here-considered.

(ii)~{\it\zb-technique:}
As for the \sz-technique, the $B$-modes angular power spectrum $C_{\ell}^B$ is reconstructed unbiased.
The dashed-dotted red curve depicts the uncertainties on $C_{\ell}^B$ via the \zb-approach using harmonic-variance optimized apodizations calculated for the \sz-approach while the colored dashed curves represent the window function with different apodization lengths $\theta_\mathrm{apo}$ ranging from 5 to 8 degrees. We have checked that using apodization length either smaller than 5 degrees or wider than 8 degrees systematically lead to higher uncertainties. For this technique, one cannot {\it a priori} apply the pixel domain computation of the variance-optimized apodizations. We nevertheless check that this is indeed the case using numerical experiments. Our results shows that weighting the maps of the Stokes parameters with the spin-0 pixel variance-optimized apodizations as derived for the \sz-technique leads to very high uncertainties for $\ell<100$.
At low multipoles, larger apodization length reduces the $E$-to-$B$ leakage lowering the uncertainties on $C_{\ell}^B$. At high multipoles, uncertainties are driven by the sky cut which raise as $\theta_\mathrm{apo}$. The harmonic-optimized window functions give the smallest uncertainties on $C_{\ell}^B$ for $\ell>100$ but, as expected, fails to provide the smallest uncertainties for $\ell<100$. For those large angular scales, the recovery of $C^B_\ell$ is only possible for $\ell>20$ and making use of analytic sky apodization.

(iii)~{\it\kn-technique:}
The estimation of the angular power spectrum appears to be biased. The solid red curve shows the estimated $C_{\ell}^B$ for an apodization length of $30'$ and is biased in the four first bins. The more we decrease the length of apodization, the less the estimated $C_{\ell}^B$ is biased to get an unbiased estimation with $\theta_\mathrm{apo}=1$ degree. 
This bias comes from the approximation $K_{\ell \ell'}^{(-)} = 0$ which is not verify in practice.
The uncertainties as derived in the \kn-approach are depicted in the lower panel of Fig.~\ref{fig:cellwmap}. Those error bars have been obtained by first computing the map of $\tilde\chi^B$ using a $C^2$ window function with an apodization length $\theta_\mathrm{apo}$ and then by removing those pixels for which the sky apodization is varying (that is an external layer with a width $\theta_\mathrm{apo}$). The three here-adopted values for $\theta_\mathrm{apo}$ are 0.5, 1 and 2 degrees. As expected from the mode counting estimation, the lowest error bars are achieved for the highest sky coverage, that is for $\theta_\mathrm{apo}=0.5$ degree. Nonetheless, for the two first bin, the error bars for the three values of $f_{\mathrm{sky}}$ are higher than the value of the signal meaning it is impossible to detect the primordial part. They decrease up to $\ell \simeq 200$ and then behave like the mode-counting uncertainty until $\ell = 1020$.

\subsubsection{Pseudo power spectrum}
\begin{figure}[hb]
\center
	\includegraphics[scale=0.52]{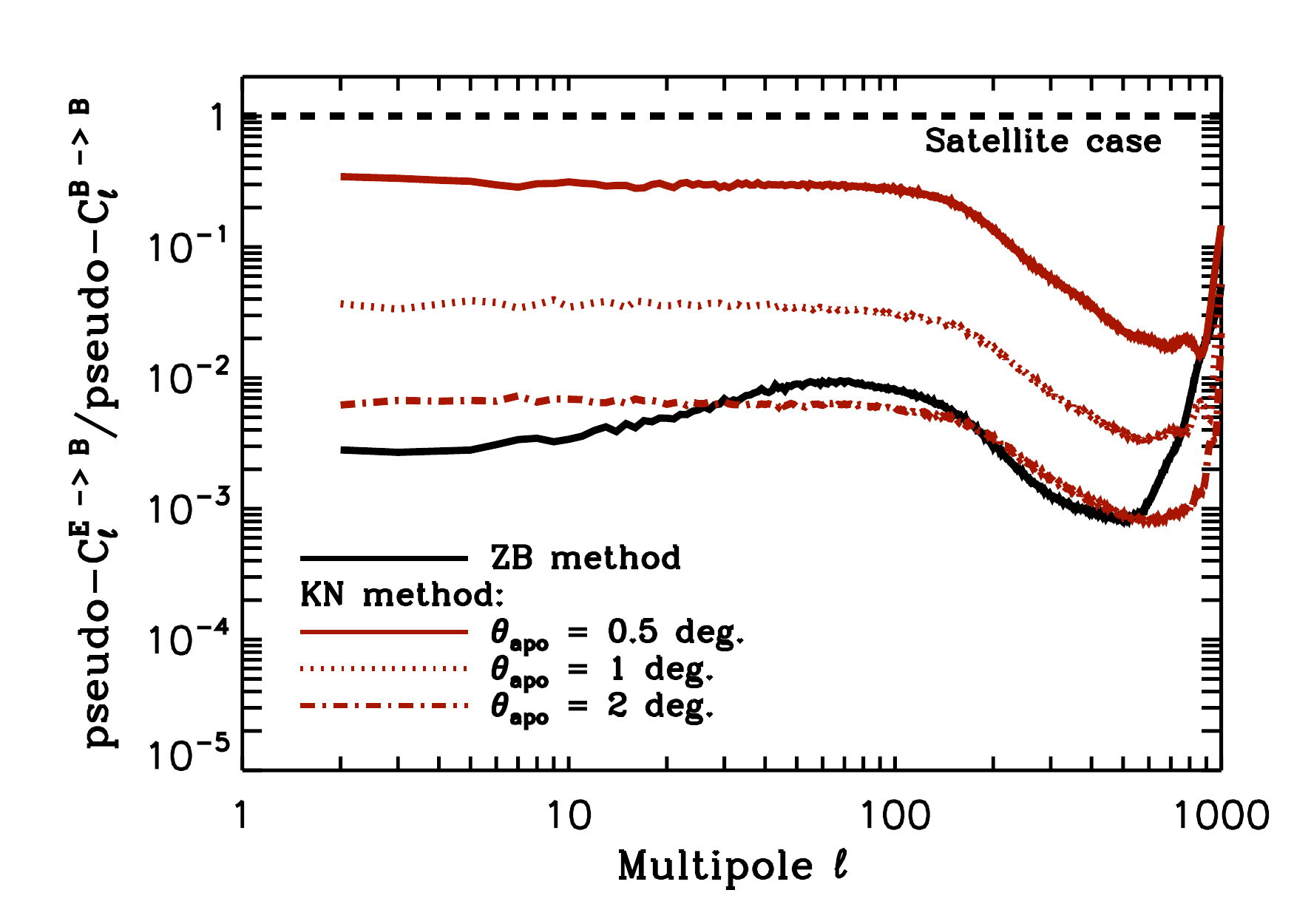}
	\caption{Ratio between $\tilde{C}_{\ell}^{E \to B}$ ($E$-mode power spectrum is derived from WMAP 7yrs best fit) and  $\tilde{C}_{\ell}^{B \to B}$ computed by correcting for such a leakage using the \zb~and \kn~$E/B$ separation techniques (respectively represented by the black curve and the red curves). This ratio amounts the leakage of $E$-modes into $B$. The dashed-black line is the benchmark with which the pseudo-$C_{\ell}$ has to be compared to. For the \kn-method, the three curves are the value of the ratio for the different values of the apodization length: $\theta_{ap} = 0.5, 1$ and $2$ degrees from top to bottom, respectively. The sky coverage is the one expected for a satellite-like experiment with holes due to point-sources removal (see upper panel of Figure \ref{fig:masks}).}
	\label{fig:pclwmap}
\end{figure}

A way of qualitatively describe potential bias in the methods is to study the $B$-modes pseudo-power spectrum $\tilde{C}_{\ell}^B$. 
Comparing these two quantities allows for a quantitative description of the leakage that bias the $B$-mode pseudo-power spectrum.
In Fig.~\ref{fig:pclwmap}, we plot the ratios $\tilde{C}_{\ell}^{E\to B}/\tilde{C}_{\ell}^{B\to B}$ for the \zb~and \kn~methods. First of all, this ratio is not zero because of the pixelization effects. This may bias the final estimate of $C^B_\ell$ if such residual leakage is not corrected for via a non-zero $K^{(-)}_{\ell\ell'}$ {\it and}  $\tilde{C}^{E\to B}_{\ell}$ cannot be safely neglected compared to $\tilde{C}^{B\to B}_{\ell}$. For the \sz-technique, those residual leakages are corrected for via the implementation of $K^{(-)}_{\ell\ell'}$. However, such an off-diagonal block of the mode-mode coupling matrices cannot be computed in the \zb- and \kn-techniques. The block $K^{(-)}_{\ell\ell'}$ is systematically set equal to zero which implicitly assumes that {\it effectively} $\tilde{C}^{E\to B}_{\ell}\ll\tilde{C}^{B\to B}_{\ell}$.
Fig.~\ref{fig:pclwmap} (the solid-black curve)indicates that this assumption is valid for the \zb-technique, the ratio being approximatively equal to $10^{-2}$ at most. On the contrary, Fig.~\ref{fig:pclwmap} (red curves) shows that $\tilde{C}^{E\to B}_{\ell}$ cannot be neglected with respect to $\tilde{C}^{B \to B}_{\ell}$ for \kn-method inducing a bias in the $B$-modes angular power spectrum as seen in the lower panel of Fig.~\ref{fig:cellwmap}.

\subsubsection{Effect of point sources in the mask}
\begin{figure}
\center
	\includegraphics[scale=0.5]{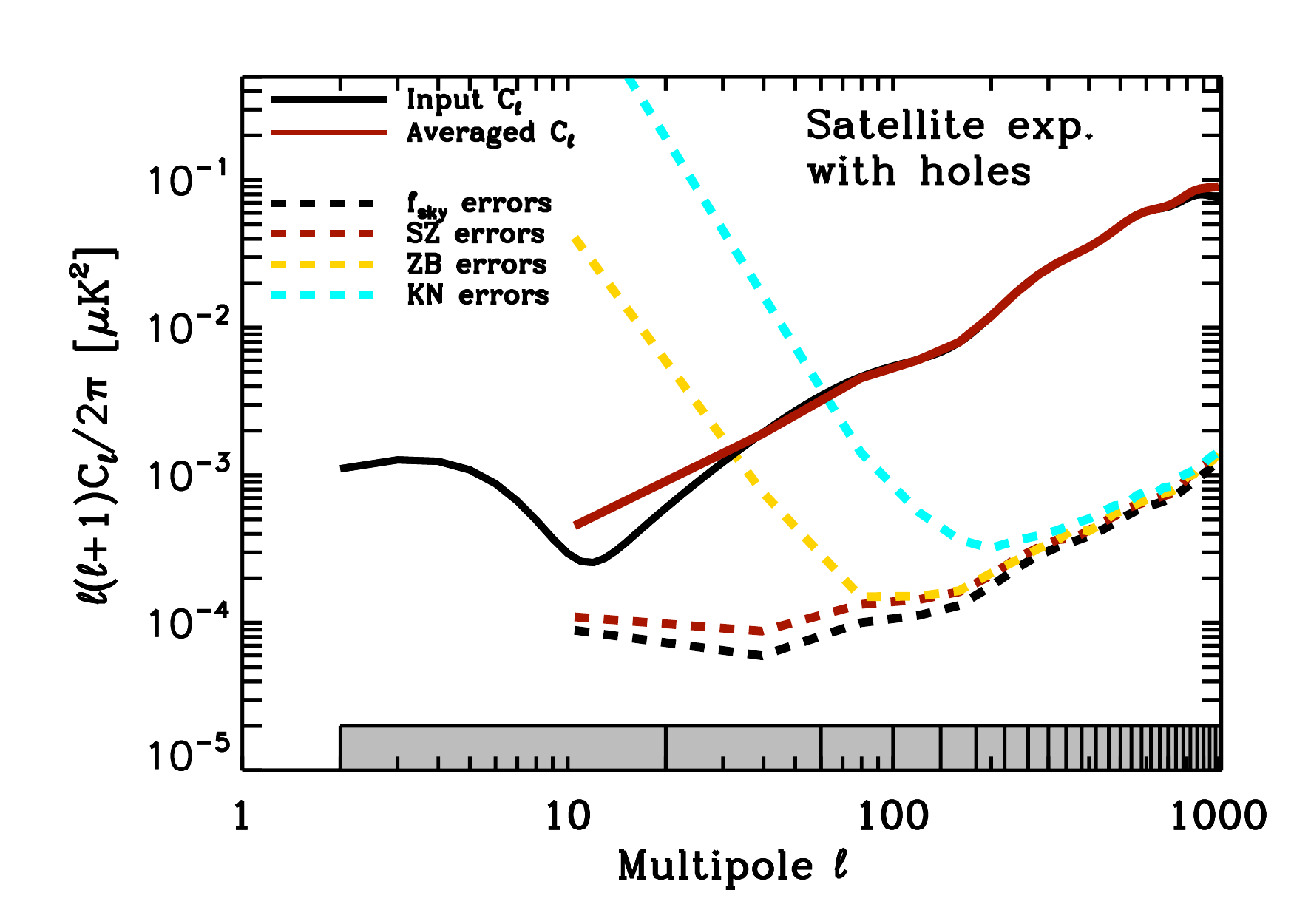} \includegraphics[scale=0.5]{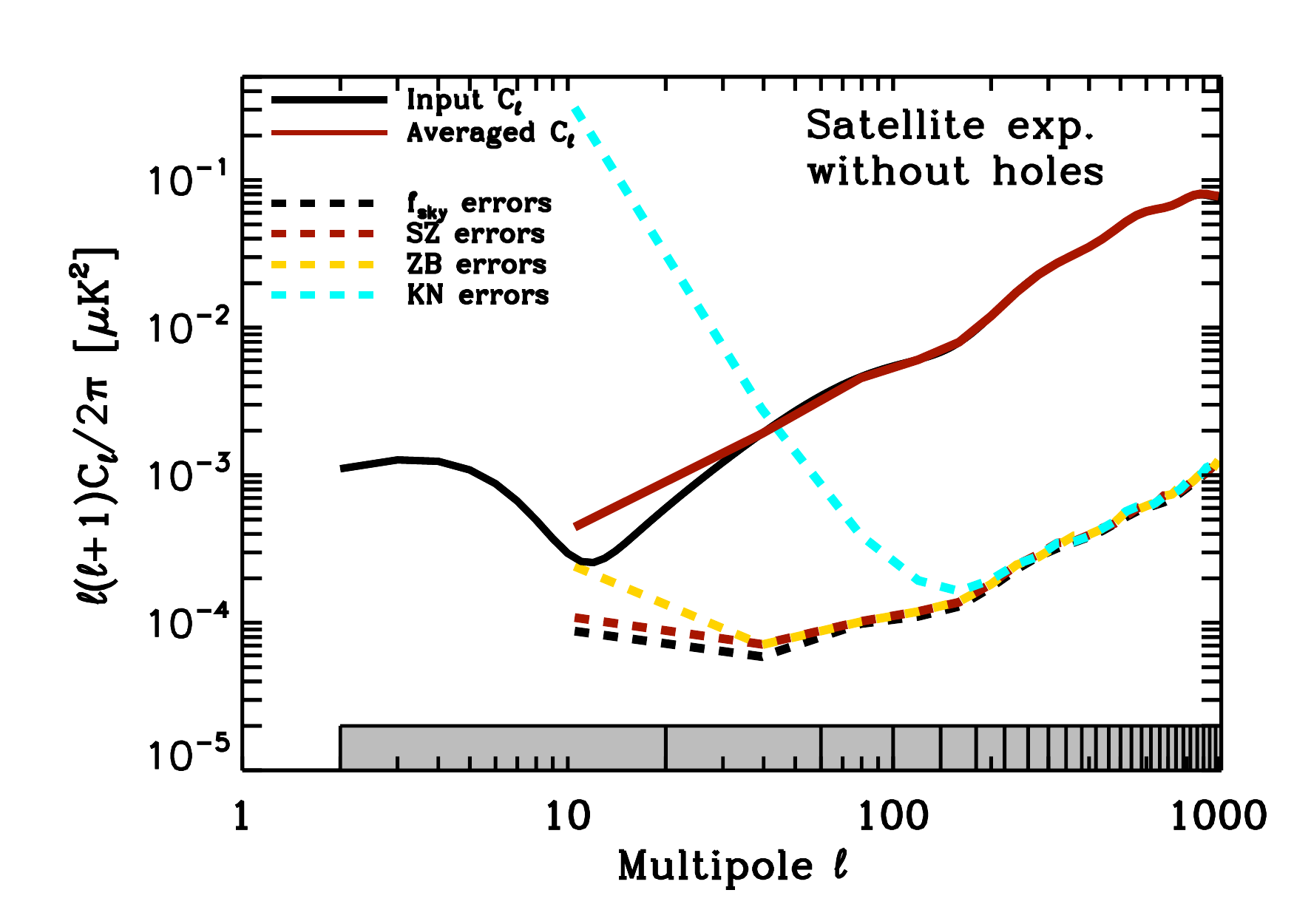}
	\caption{Power spectrum uncertainties on $B$-modes using cross-spectrum estimation for the case of satellite experiment with holes mimicking point sources-removal ($f_\mathrm{sky}\sim71\%$). The red dashed line represents the variance obtained via \sz-method, the blue dashed is via \zb-method and at last the yellow dashed dotted for the \kn-method.}
	\label{fig:sumwmapnoholes}
\end{figure}

Furthermore, as already highlighted in Sec. \ref{sssec:pcg}, we confirm the importance of the point sources holes in the mask. Indeed, we also calculated the $B$-modes angular power spectra for a mask which do not account for the polarized point sources ($f_\mathrm{sky} = 73\%$). The lowest achieved uncertainties for each method are depicted on Fig. \ref{fig:sumwmapnoholes} with holes (upper panel) and without holes (lower panel). The difference between the two $f_\mathrm{sky}$ being $2\%$, one could expect from a na\"\i{ve} mode counting, the error bars to increase by a factor $\sim1.01$ by adding holes. Though such a scaling indeed applies to the case of the \sz-method, it appears that both the \zb- and the \kn-method are very sensitive to the presence of holes at large angular scales. Clearly, the uncertainties increase by more than $\sim1.01$ by adding holes for $\ell<140$ for both the \zb- and \kn-methods. Though the \sz-technique can handle the impact of holes, the increase of the variance at large scales for the \zb- and \kn-techniques shows that a dedicated treatment of holes could be mandatory.

It is instructive to compare the \sz-approach to the \zb-approach to understand why the latter can deals with holes while the former don't. They differ one from each other by the use of two different sky apodizations ; the pixel-domain, variance-optimized apodization for the \sz-technique and the harmonic-domain, variance-optimized sky apodization for the \zb-technique. If one uses the harmonic-domain, variance-optimized sky apodization, the \sz-approach would suffer from the high increase of the variance at large angular scales similar to the increase of the variance observed in the \zb-approach. In other words, all the additional complexity due to holes in the mask is nicely treated in the \sz-approach thanks to its flexibility and a dedicated computation of the sky apodization in the pixel domain.

\subsubsection{Conclusion for satellite-like experiment}
To summarize, the \sz-method gives unbiased $B$-mode power spectra and the smallest uncertainties, close to the mode-counting one, for the case of a large sky coverage (see Fig. \ref{fig:varsz} for a reconstruction multipole by multipole and Fig.~\ref{fig:sumwmapnoholes} for a reconstruction within bandpower). The results with the \zb-method with the harmonic-optimized windows are similar to those of the \sz-method for $\ell>100$. For $\ell\in[20,100]$, estimating $C^B_\ell$ is still possible but with a smaller significance. For $\ell<20$, the \zb-method fails to reconstruct the $B$-modes angular power spectra. Our implementation of the \kn-method does not manage to reconstruct an unbiased $C_{\ell}^B$ for the four first bins if the apodization length is too small. For those apodization allowing the \kn-method to provide an unbiased estimation ($\theta_\mathrm{apo}\geq1$ degree), reconstructing $C^B_\ell$ is not possible for $\ell<60$. For intermediate angular scales, $60<\ell<300$, the reconstruction is possible with a lower signal-to-noise ratio than the one achieved thanks to either the \sz-technique or the \zb-technique.


\section{Results: small scale experiment}

\begin{figure}
\center
	\includegraphics[scale=0.5]{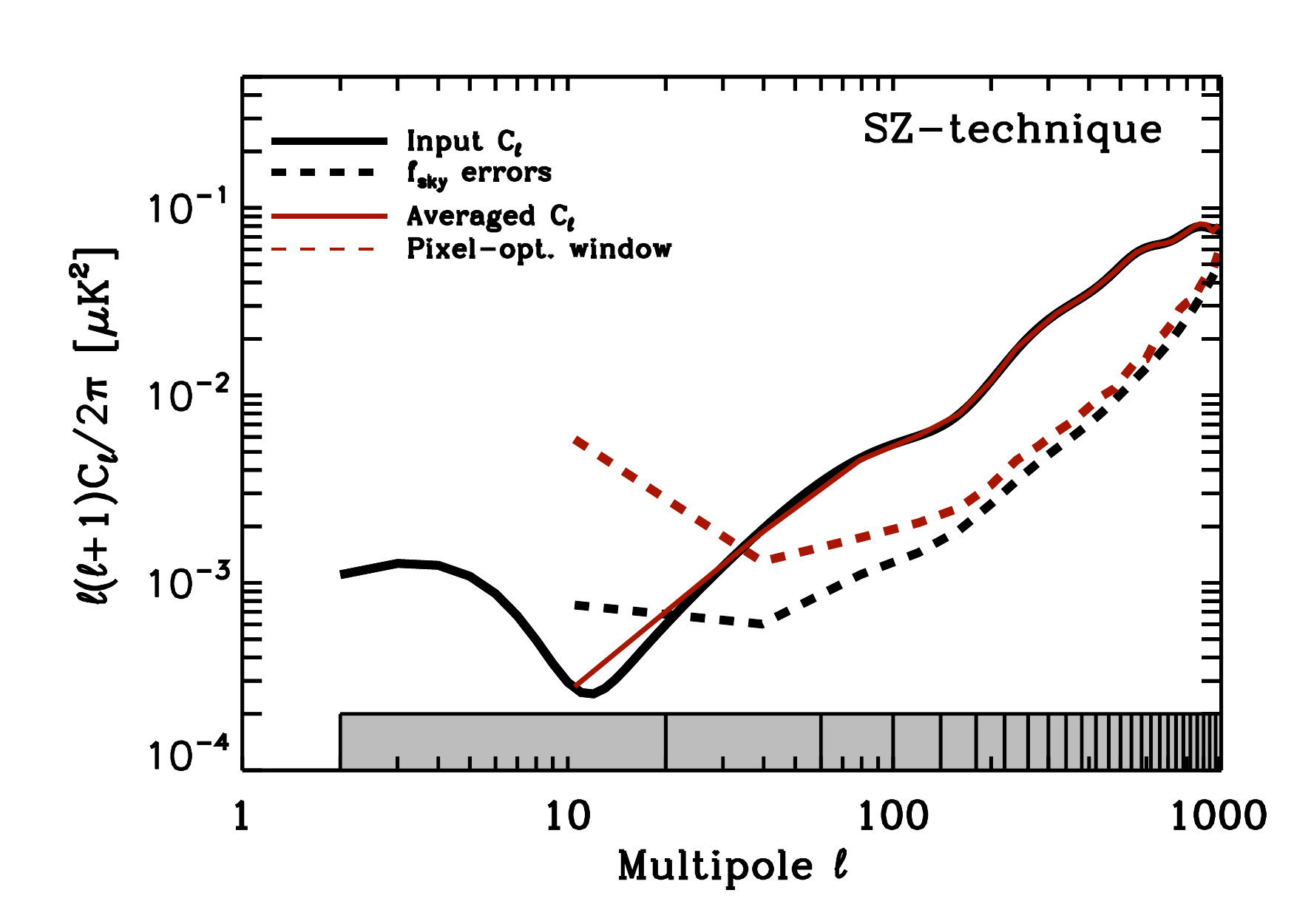} \\
	\includegraphics[scale=0.5]{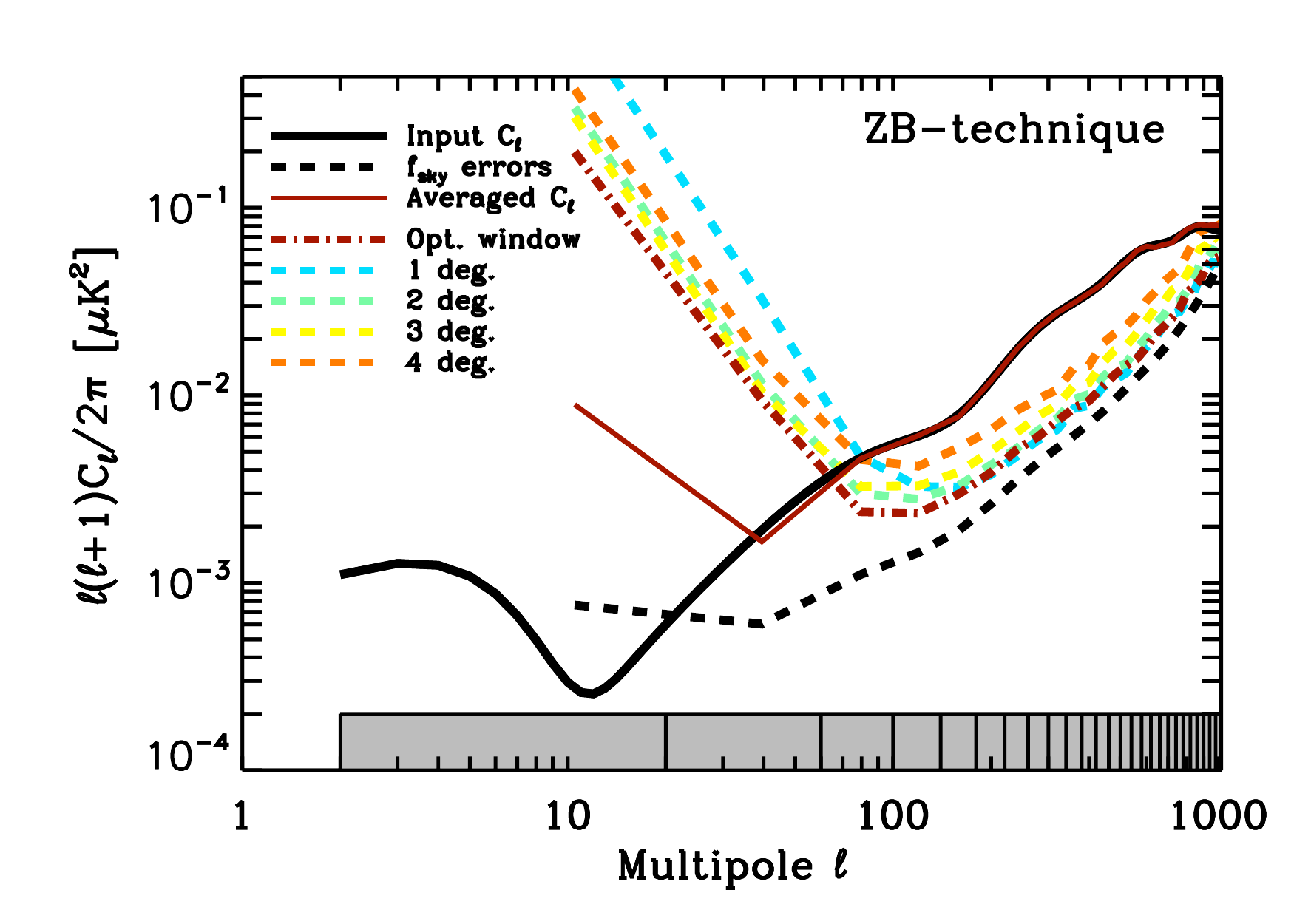} \\
	\includegraphics[scale=0.5]{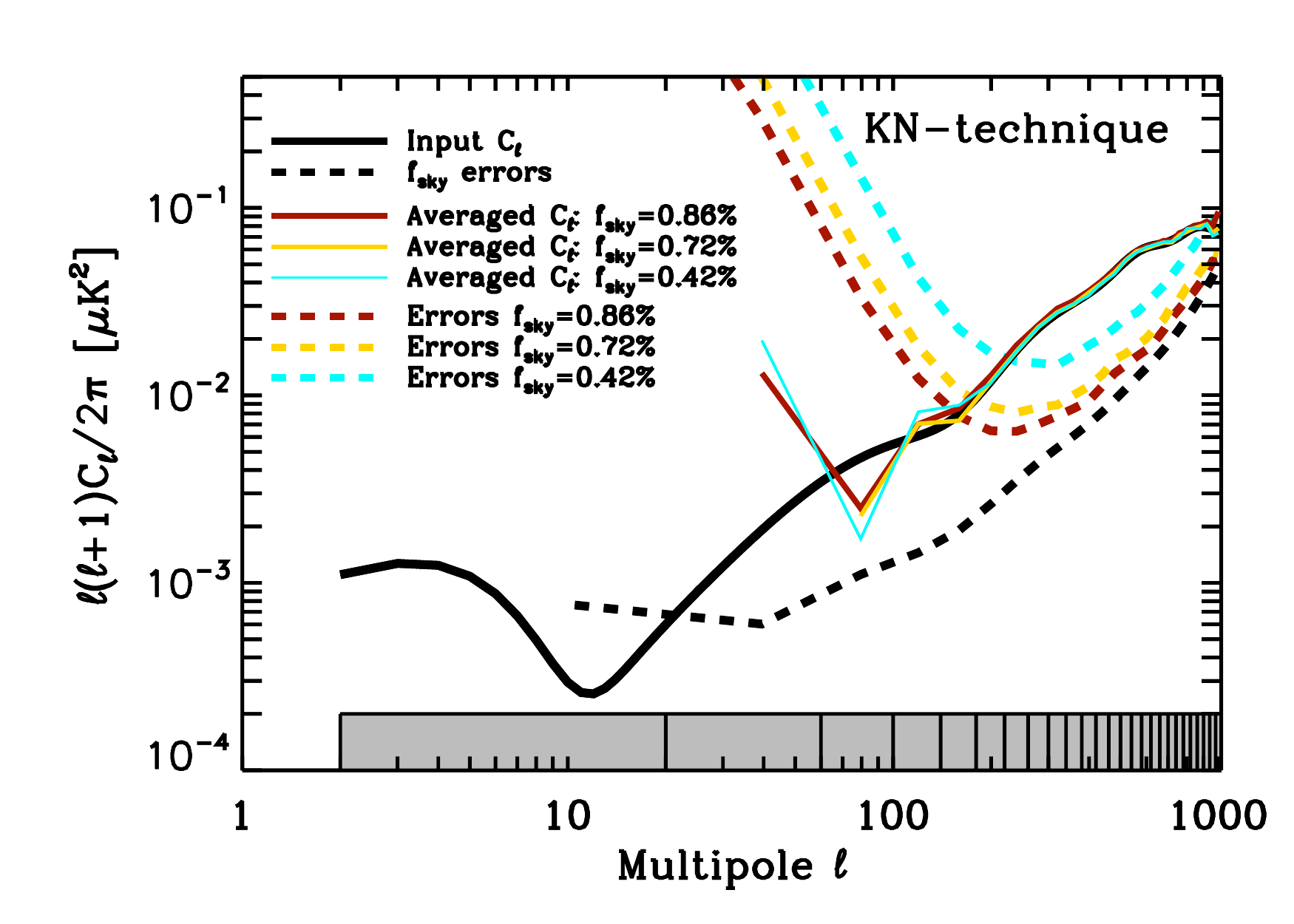}
	\caption{Power spectrum uncertainties on $B$-modes using cross-spectrum estimation for the case of balloon-borne experiment ($f_\mathrm{sky}\sim1\%$). Upper, middle and lower panels are respectively for the \sz-, \zb- and \kn-methods. The sky apodizations used for each techniques are described in Sec.~\ref{sect:formalism}}
	\label{fig:cellebex}
\end{figure}

In the case of a balloon-born like experiment, the reconstructed $B$-mode angular power spectra and their associated uncertainties are shown in Fig. \ref{fig:cellebex} for the three techniques. Those angular power spectra are estimated from $\ell=2$ to $\ell=1020$ with the first bin ranging from 2 to 20 and the following bins having a bandwidth $\Delta \ell$ equal to 40. We underline that for such a small value of the sky coverage, the amplitude of the binned $C^B_\ell$ in the first bin $b_1=[2,20]$, is $C^B_{b_1}\simeq5.9\times10^{-4}\mu K^2$ for $r=0.05$. The Fisher estimate of the uncertainties for the same value of $r$ leads to $\sqrt{\mathbf{\Sigma}_{b_1b_1}}\simeq7.6\times10^{-4}\mu K^2$. Detecting a non-vanishing $C^B_\ell$ at angular scales between $\ell=2$ and $\ell=20$ appears unfeasible for small-scale experiments since the Fisher calculation underestimates the variance on the pseudo-$C_\ell$ reconstruction of angular power spectra. On each of the three graphs, the solid-black curve corresponds to the input $B$-mode power spectrum to be estimated while the solid-red curve stands for the estimated angular power spectrum averaged over 500 simulations. The dashed-black curves correspond to the mode-counting estimate of power spectrum uncertainties obtained with $f_\mathrm{sky}=1\%$ which serves as a benchmark. For each of the graphs, the dashed-colored curves stands for MC estimations of the power spectrum uncertainties for each of the techniques.

(i)~{\it\sz-technique:}
We confirm that the reconstructed angular $B$-mode power spectrum is unbiased for the entire range of multipoles considered here.
As previously mentioned, we only use pixel-optimized window functions for the case of the \sz-technique (upper panel of Fig. \ref{fig:cellebex}) and the displayed error bars are therefore the lowest ones to be expected in such an approach. We refer the reader to \cite{grain_etal_2009} for an exhaustive discussion on the performances of such a technique. The relevant conclusion in such a case is that a precise enough estimation of $C^B_\ell$ is achieved for multipoles starting from $\ell = 20$ to $\ell = 1020$.

(ii)~{\it\zb-technique:}
In such a case, the estimated $C^B_\ell$'s are also unbiased from $\ell=2$ to $\ell=1020$.
We show the power spectrum uncertainties for two kind of windowing. Dashed-colored curves ranging from blue to orange stand for error bars derived using a $C^2$ window function with an apodization length varying from 1 degree to 4 degrees. It clearly shows that depending on angular scales, the apodization length has to be adapted to reach the lowest uncertainties. For the three first bins, {\it i.e.} $2\leq\ell<100$, an apodization length of 3 degrees provides the lowest error bars. For higher multipoles, an apodization length of 1 degree leads to the smallest error bars. The dashed-red curve corresponds to the uncertainties on the reconstructed $C^B_\ell$'s using optimized window function computed in harmonic domain\footnote{Because the contour of the mask are rather simple for the small-scale experiment, the harmonic computation of the variance-optimized apodizations leads to very similar results to the pixel domain computation for the \sz-technique.}. This clearly shows that unlike the case of a satellite mission, using such harmonic variance-optimized sky apodizations provides the lowest error bars on the entire angular range. However, though very efficient at multipoles greater than 60, this approach fails to reconstruct the $B$-mode angular power spectrum for the two first bins comprised in $2 \leq\ell<20$ and $20\leq\ell<60$.

(iii)~{\it\kn-technique:}
The \kn-technique provides an unbiased $B$-modes angular power spectrum though highly scattered because of the high level of the variance at low $\ell$.
As for the discussed case of a satellite-like experiment, the lowest uncertainties are obtained for the highest sky coverage \textit{i.e.} for $\theta_\mathrm{apo} = 0.5$ degrees though the \kn-technique is able to estimate $C^B_\ell$ only for $\ell$ values greater than $\sim150$ and therefore 'misses' the bump at $\ell\sim100$ due to the primordial component of the $B$-mode angular power spectrum.

\begin{figure}
\center
	\includegraphics[scale=0.5]{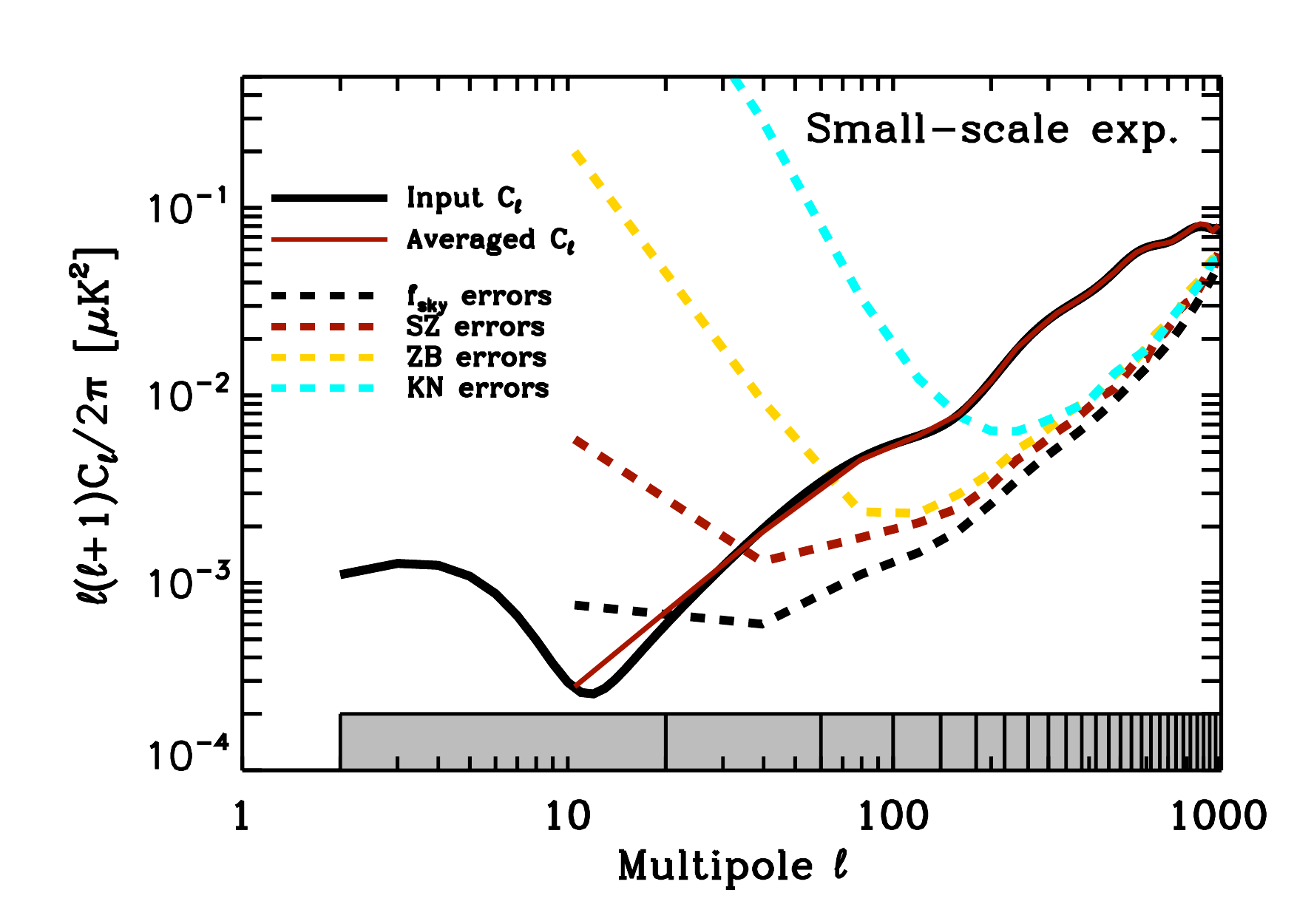} 
	\caption{Power spectrum uncertainties for each of the three techniques for the case of a small-scale experiment with $f_\mathrm{sky}\simeq1\%$, $\sigma_Q=5.75~\mu K$-arc minute and $\theta_{\mathrm{Beam}}=8~$arc minutes. Dashed-red, dashed-cyan and dashed-yellow curves are respectively for the \sz-, \zb- and \kn-techniques. The dashed-black curve stand for $f_\mathrm{sky}$ estimate of the error bars.}
	\label{fig:cellsumebex}
\end{figure}

Figure \ref{fig:cellsumebex} summarizes our results depicting the lowest error bars on the $B$-mode estimation for each of the three techniques. From those results, it is rather obvious that the \sz-technique performs the best for power spectrum reconstruction from both the viewpoint of bias and uncertainties. This approach allows for an accurate enough estimation of $C^B_\ell$ for $\ell\geq20$ while the \zb-technique and the \kn-technique allow for such a reconstruction for $\ell\geq60$ and $\ell\geq150$ respectively. Those differences may drastically affect our ability to set constraints on those cosmological parameters probing the inflationary phase as {\it e.g.} the tensor-to-scalar ratio $r$. We remind that the primordial component of $C^B_\ell$ --from which constraint on $r$ can be set-- is dominant for $\ell$ values lower or equal to $\sim100$ while the lensing-induced $B$-modes start to dominate the angular power spectrum for $\ell>100$. With our binning, this means that with the \sz-technique, one can detect the primordial $B$-modes in two bins ({\it i.e.} $\ell\in[20,60]$ and $\ell\in[60,100]$). 
With the \zb-technique, the primordial component of $C^B_\ell$ can be detected in only one bin, $\ell\in[60,100]$, while a detection of the primordial component seems impossible with the \kn-approach\footnote{Strictly speaking, some constraint can be set on $r$ even by using the \kn-approach (at least some upper limit). But this may probably prevent for any {\it measurement} of $r$.}.

\section{Conclusion and discussion}
\label{sect:conclusion}
We first presented three different pseudo-spectrum estimators designed to remove -or at least reduce- the $E$-to-$B$ leakage which may compromise any detection of the $B$-modes and especially its primordial part. We then test the relative efficiency of those estimators to reconstruct the $B$-modes angular power spectrum through Monte-Carlo. Two different kinds of sky coverage have been chosen for our analysis: a small scale coverage (observed part of the sky $\sim 1\%$) and a large coverage of the celestial sphere as motivated by a future satellite mission dedicated to $B$-modes detection with $f_\mathrm{sky} \sim 71\%$. Both sky-coverage incorporates holes mimicking point-sources removal.

All  three techniques studied here try to reconstruct, implicitly or explicitly, the $\chi^B$ field, which is known to contain only the $B$-modes. We first described the so-called \sz-method which efficiency lies in an adapted choice of basis to decompose the $E$- and $B$-modes optimizing the apodization of the applied mask. Then, the \zb-technique principle is developed. It consists in calculating the masked $\chi^B$ with an adapted apodized mask, it implies derivation operations of the masked polarization field which are actually done in the harmonic space. Finally the \kn-method is based on the fact that applying a mask on the reconstructed $B$-modes reduces significantly the level of $E$-to-$B$ leakage. In this article, we do not claim to exactly implement the methods as they were described in the referred articles. Slight changes have been made in their implementation in order to minimize as much as possible the effective $E$-to-$B$ leakage. 

We compare the results of those methods on each of our simulation set.

First, we found that correcting for $E$-to-$B$ leakages at both the levels of mean and variance is mandatory in the case of a satellite mission covering $\sim71\%$ of the sky for an efficient recovery of the primordial component of $B$-modes, $\ell<30$. Moreover, we have shown that the intricate shape of the galactic mask makes the uncertainties of the reconstructed $C^B_\ell$ using methods correcting for $E$-to-$B$ leakages, very sensitive to sky apodization applied to $Q$ and $U$ maps for $\ell<60$. An efficient computation of variance-optimized sky apodization is therefore crucial for the applicability of those methods. From that practical perspective, the \sz-method appears to better armed as it offers some flexibility in the computation of the sky apodization. 

Second, we computed the pseudo-$C_{\ell}$ which amounts the $E$-modes leaking into $B$ applying the three different techniques. Each of them are able to significantly decrease the $E$-to-$B$ leakage though none manage to exactly cancel it because of the pixelization effects. Nonetheless, the value of the uncertainties on the $C_{\ell}^B$ reconstruction is the key issue because it tells us if a detection is possible or not. As shown by our numerical results, the final uncertainties on the estimated $B$-modes power spectra can overwhelm the signal even when the $E$-to-$B$ leakage is well controlled. The \sz-method gives the smallest error bars on the $B$-modes angular power spectra for both the large and small scale experiments as they follow quite well the mode-counting uncertainties. Though we can not recover the largest angular scales $\ell$-by-$\ell$ for $\ell \le 5$ with the \sz-approach, we can reach a detection for such scales using the appropriate binning. 
The \zb-method, as explained in Sec. \ref{sec:comparison}, is \textit{theoretically} equivalent to the \sz~one. From the numerical results, we showed that \textit{practically} this method is less efficient at large angular scales ($\ell<60$ for a satellite-like mission and $\ell<100$ for a small sky survey), allowing us to reconstruct $C^B_\ell$ starting at $\ell\sim20$ for a satellite-like mission (resp. starting at $\ell=60$ for a small sky survey). For smaller angular scales $\ell>60$, these two methods provide similar results. The \kn-method is by construction expected to be less efficient than the other methods in our implementation, as described in \ref{sec:comparison}. Indeed, the sky coverage reduces according to the apodization length leading to a higher variance compared to \sz~estimator. The power spectrum analysis shows that this method is reliable for high $\ell$, but the error bars overwhelm the signal for the two first bins ({\it i.e.} $\ell<60$) in the case of a fraction of the sky of $71\%$ (resp. for the three first bins, $\ell<100$, for a small sky survey).

The two figures \ref{fig:cellsumebex} and \ref{fig:sumwmapnoholes} sum up the errors made on the estimated $C_{\ell}^B$ via the \sz-, \zb-~and \kn-methods in the two experimental configurations. In the way we have implemented those techniques, the \sz-method is the most efficient one. For both type of experimental set-ups, it makes possible the estimation of $C^B_\ell$ with uncertainties on par with the most optimistic, Fisher estimates. The key step making the \sz-approach more efficient is its flexibility in terms of sky apodization. This is highlighted in Fig. \ref{fig:varsz}: relaxing the derivative relationship relating the spin-1 and spin-2 windows to the spin-0 window is mandatory for computing variance-optimized sky apodizations, drastically lowering the final uncertainties on the estimated $C^B_\ell$'s. However, neither the \zb- nor the \kn-approaches are currently designed to offer such a flexibility. We have checked that if one uses the same sky apodization (for example an analytic window function with a given apodization length) the \sz- and \zb-methods leads to similar uncertainties. Inversely, we have also checked that one cannot use the pixel-domain, variance-optimized sky apodization (relaxing the derivative relationship) in the \zb-approach as it systematically leads to an increase of the final error bars as compared to {\it e.g.} using analytic windows with an appropriate choice of the apodization length. This shows that the in applicability of those pseudo-$C_\ell$ estimators which do not mix $E$ and $B$-modes is highly conditionned by the pre-computation of variance-optimized sky apodization.

\begin{acknowledgments}
This research used resources of the National Energy Research Scientific Computing Center, which is supported by the Office of Science of the U.S. Department of Energy under Contract No. DE-AC02-05CH11231. Some of the results in this paper have been derived using {\sc s$^2$hat} \cite{s2hat,pures2hat,hupca_etal_2010,szydlarski_etal_2011}, {\sc healpix}~\cite{gorski_etal_2005} and {\sc camb}~\cite{lewis_etal_2000} software packages.\end{acknowledgments}


\appendix
\section{Convolution kernels for the \kn-method}
\label{app:knkernel}
We show in this appendix that a complete derivation of the convolution kernels in the \kn-approach is computationally prohibitive. 

In such an approach, a map of the masked $\chi^B$ field is first built by applying Eq. (\ref{eq:chipixel}). As a function of the 'true' CMB $E$- and $B$-multipoles, this masked $\chi^B$ field reads:
\begin{equation}
	\tilde{\chi}^B(\vec{n})=-\ds\sum_{\ell'm'}\left[\mathcal{K}^{(-)}_{\ell'm'}(\vec{n})a^E_{\ell'm'}+i\mathcal{K}^{(+)}_{\ell'm'}(\vec{n})a^B_{\ell'm'}\right].
\end{equation}
The above convolution kernels should be viewed as scalar functions in the pixel domain parametrized by some harmonic indices. They measure the amount of $(\ell,m)$-multipoles of $B(E)$ types contributing to the {\it masked} $\chi^B$ field at direction $\vec{n}$. In principle, those coupling functions are given by :
\begin{equation}
\mathcal{K}^{(\pm)}_{\ell m}(\vec{n})=\lambda_{\pm}\left[\bar\partial\bar\partial\left(M\,{}_{2}Y_{\ell m}\right)\pm\partial\partial\left(M\,{}_{-2}Y_{\ell m}\right)\right],
\end{equation}
with $\lambda_{\pm}$ a complex valued numerical constant. Expanding the spin-raising and spin-lowering operation, we obtain
\begin{eqnarray}
	\mathcal{K}^{(+)}_{\ell m}(\vec{n})&=&N_{\ell,2}\times M\times Y_{\ell m} \\
	&+&\lambda_{+,1,\ell}\left[{}_{1}Y_{\ell m}\bar\partial{M}+{}_{-1}Y_{\ell m}\partial{M}\right] \nonumber \\
	&+&\lambda_{+,2,\ell}\left[{}_{2}Y_{\ell m}\bar\partial\bar\partial{M}+{}_{-2}Y_{\ell m}\partial\partial{M}\right] \nonumber
\end{eqnarray}
and
\begin{eqnarray}
	\mathcal{K}^{(-)}_{\ell m}(\vec{n})&=&\lambda_{-,1,\ell}\left[{}_{1}Y_{\ell m}\bar\partial{M}-{}_{-1}Y_{\ell m}\partial{M}\right] \\
	&+&\lambda_{-,2,\ell}\left[{}_{2}Y_{\ell m}\bar\partial\bar\partial{M}-{}_{-2}Y_{\ell m}\partial\partial{M}\right], \nonumber
\end{eqnarray}
where the explicit expression of the $\lambda_i$'s are of no importance here. It is clear from the above computation that where the mask is constant, i) $\mathcal{K}^{(-)}_{\ell m}(\vec{n})$ is zero and there is no $E$-to-$B$ leakage and ii) $\tilde{\chi}^B=\sum_{\ell m}N_{\ell,2}\times M\times Y_{\ell m}$ which is just the definition of the $\chi^B$ field on the mask $M$. 

Here, we just re-confirmed that the derivation of $\chi^B$ proposed in \cite{kim_naselsky_2010} is exact {\it on the part of the sky} where the mask is {\it constant}. The above result is made possible if and only if the convolution kernels $F_{\pm}$ computed in the harmonic domain (see Eq. \eref{kn:kernel1}) is effectively a precise enough representation of the operator $\left[\partial\partial  \pm \bar\partial\bar\partial \right]$. However, the truncation in the $(\ell,m)$ summation and pixelization shows that it is not the case. Indeed, if it was the case, the $F_{\pm}(\vec{n},\vec{n}')$ would be completely local and the map of the leaking $E$-modes would be concentrated on the boundaries of the observed sky. But the results displayed in \cite{kim_naselsky_2010} shows that $F_\pm$ is not local, though well-peaked, and that leaked $E$-modes extends inside the observed patch. As a consequence, the $\mathcal{K}^{(\pm)}_{\ell m}$ functions are not strictly equal to the above expressions leading to residual $E$-to-$B$ leakages as well as potential $\ell$-to-$\ell'$ aliasing. Those functions should be computed differently in order to keep track of, at least, the $(\ell,m)$ truncation and subsequently derived an {\it unbiased} pseudo-$C_\ell$ estimator by correcting for the different residual leakages. For this purpose, we propose here an alternative expression for the $\mathcal{K}^{(-)}_{\ell m}(\vec{n})$ which can then be plugged in the final expression of the pseudo-$C_\ell$ estimator.

The $(\ell,m)$ to pixel convolution kernels are expressed as functions of Wigner-$3j$ symbols and the multipoles of the binary mask, $M$, describing the observed sky:
\begin{eqnarray}
	\mathcal{K}^{(\pm)}_{\ell'm'}(\vec{n})&=&\frac{i}{2}\ds\sum_{\ell_1m_1}\sum_{\ell_2m_2}(-1)^{m_1}N_{\ell_1,2}F(\ell',\ell_1,\ell_2) \\
		&\times&\left(\begin{array}{ccc}
			\ell' & \ell_1 & \ell_2 \\
			m' & -m_1 & m_2
		\end{array}\right)\mathcal{M}_{\ell_2m_2}Y_{\ell_1m_1}(\vec{n}) \nonumber \\
		&\times&\left[\left(\begin{array}{ccc}
			\ell' & \ell_1 & \ell_2 \\
			2 & -2 & 0
		\end{array}\right)\pm\left(\begin{array}{ccc}
			\ell' & \ell_1 & \ell_2 \\
			-2 & 2 & 0
		\end{array}\right)\right], \nonumber
\end{eqnarray}
with
\begin{equation}
	F(\ell,\ell',\ell'')=\sqrt{\frac{(2\ell+1)(2\ell'+1)(2\ell''+1)}{4\pi}}.
\end{equation}
Being scalar functions, there multipoles are obtained by projecting them on the spherical harmonic basis and read
\begin{eqnarray}
	\mathcal{K}^{(\pm)}_{\ell' m'}(\ell,m)&=&\ds\int \mathcal{K}^{(\pm)}_{\ell'm'}(\vec{n})Y^\star_{\ell m'}(\vec{n}) \\
		&=&\frac{i}{2}\ds\sum_{\ell_2m_2}(-1)^{m}N_{\ell_1,2}F(\ell,\ell',\ell_2) \nonumber \\
		&\times&\left(\begin{array}{ccc}
			\ell' & \ell & \ell_2 \\
			m' & -m & m_2
		\end{array}\right)\mathcal{M}_{\ell_2m_2} \nonumber \\
		&\times&\left[\left(\begin{array}{ccc}
			\ell' & \ell & \ell_2 \\
			2 & -2 & 0
		\end{array}\right)\pm\left(\begin{array}{ccc}
			\ell' & \ell & \ell_2 \\
			-2 & 2 & 0
		\end{array}\right)\right]. \nonumber
\end{eqnarray}
Secondly, the reconstructed $\tilde{\chi}^B$ field is masked again with the $M_{\chi^B}$ from which pseudo-multipoles, denoted $\tilde{\chi}^B_{\ell m}$ hereafter, are derived. It is easily shown that
\begin{equation}
	\tilde{\chi}^B_{\ell m}=\ds\sum_{\ell'm'}\left[K^{(+)}_{\ell m,\ell'm'}a^B_{\ell' m'}-iK^{(-)}_{\ell m,\ell'm'}a^E_{\ell' m'}\right],
\end{equation}
with	
\begin{eqnarray}
	K^{(\pm)}_{\ell m,\ell'm'}&=&(-i)\ds\sum_{\ell_1m_1}\sum_{\ell_4m_4}(-1)^{m}F(\ell,\ell_1,\ell_3) \\
	&\times&\mathcal{K}^{(\pm)}_{\ell'm'}(\ell_1,m_1)\mathcal{M}^{(\chi^B)}_{\ell_3m_3} \nonumber \\
	&\times&\left(\begin{array}{ccc}
			\ell & \ell_1 & \ell_3 \\
			-m & m_1 & m_3
		\end{array}\right)\left(\begin{array}{ccc}
			\ell & \ell_1 & \ell_3 \\
			0 & 0 & 0
		\end{array}\right). \nonumber
\end{eqnarray}

Finally, at the level of power spectra, the following convolution kernels are in principle derived using 
\begin{eqnarray*}
	K^{(\pm)}_{\ell\ell'}=\frac{1}{2\ell+1}\sum_{m,m'}\left|K^{(\pm)}_{\ell m,\ell'm'}\right|^2.
\end{eqnarray*}
In more convential approach, the above azimuthal averaging is done analytically and allows us to greatly simplifies the expression of $K^{(\pm)}_{\ell\ell'}$. However, it is easily understood by first plugging the expression of $\mathcal{K}^{(\pm)}_{\ell'm'}(\ell,m)$ into $K^{(\pm)}_{\ell m, \ell'm'}$ and second, by plugging the expression of $K^{(\pm)}_{\ell m, \ell'm'}$ into $K^{(\pm)}_{\ell\ell'}$, that such simplifications cannot be applied in the \kn-approach. As a consequence, the computation of $K^{(\pm)}_{\ell m, \ell'm'}$ implies three summations over $(\ell,m)$ indices and the intricate multiplication of four Wigner-$3j$ symbols. It is therefore obvious that the complete derivation of the convolution kernels in the \kn-technique cannot be performed numerically. 


\section{Comparing the \kn- and \zb-method}
\label{app:knzb}
We show in this appendix that the \kn-method {\it approximates} the \zb-technique if $W$ satisfies the Dirichlet and Neuman boundary conditions. Our starting point is the first term of the rhs of Eq. (\ref{eq:chiB}):
\begin{eqnarray}
	\mathcal{B}(\vec{n})&=&\frac{i}{2}\left[\bar\partial\bar\partial\left(W(\vec{n})P_{2}(\vec{n})\right)-\partial\partial\left(W(\vec{n})P_{-2}(\vec{n})\right)\right] \\
	&=&\frac{i}{2}\ds\int d\vec{n}'\sum_{\ell m}Y_{\ell m}(\vec{n})Y^\star_{\ell m}(\vec{n}')\bigg[\bar\partial\bar\partial\left(W(\vec{n}')P_{2}(\vec{n}')\right) \nonumber \\
	&-&\partial\partial\left(W(\vec{n}')P_{-2}(\vec{n}')\right)\bigg]. \nonumber
\end{eqnarray}
The second line is obtained by inserting the closure properties of the spherical harmonics. By performing two integration by parts and using the boundary conditions verified by $W$ to cancel the contour integrals, one obtains:
\begin{eqnarray}
	\mathcal{B}(\vec{n})&=&\frac{i}{2}\ds\int d\vec{n}'\sum_{\ell m}Y_{\ell m}(\vec{n})\bigg[W(\vec{n}')P_{2}(\vec{n}')\bar\partial\bar\partial Y^\star_{\ell m}(\vec{n}')
	\nonumber
	\\
	&-&W(\vec{n}')P_{-2}(\vec{n}')\partial\partial Y^\star_{\ell m}(\vec{n}')\bigg]. \label{eq:appA}
\end{eqnarray}
We remind that 
\begin{eqnarray}
	\bar\partial\bar\partial Y^\star_{\ell m}=N_{\ell,2}\times{}_{+2}Y^\star_{\ell m}, && \partial\partial Y^\star_{\ell m}=N_{\ell,2}\times{}_{-2}Y^\star_{\ell m}. \nonumber
\end{eqnarray}
By inserting the above expression in Eq. (\ref{eq:appA}), one easily recognize the convolution kernels used in the \kn-method to finally get
\begin{eqnarray}
	\mathcal{B}(\vec{n})&=&\ds\int d\vec{n}'W(\vec{n}')\left[F_+(\vec{n},\vec{n}')P_{2}(\vec{n}')\right. \label{eq:chipixel} \\
	&-&\left.F_-(\vec{n},\vec{n}')P_{-2}(\vec{n}')\right] \nonumber
\end{eqnarray}
with
\begin{equation}
	F_{\pm}(\vec{n},\vec{n}')=\frac{i}{2}\displaystyle\sum_{\ell m}N_{\ell,2}\times Y_{\ell m}(\vec{n})\times {}_{\pm2}Y^\star_{\ell m}(\vec{n}').
	\label{kn:kernel1}
\end{equation}
This finishes our proof that the \kn-method applied to $W\times P_{\pm2}$ is equal to the first term of the central equation, {\it i.e.} Eq. (\ref{eq:chiB}), of the \zb-approach, as Eqs. (\ref{eq:chipixel}) \& (\ref{kn:kernel1}) are exactly the numerical starting point of the \kn-method (see equations (11) \& (12) of \cite{kim_naselsky_2010}).


\section{Noise bias}
\label{app:noise}
We provide in this appendix the explicit calculation of the noise bias for the \zb- and the \kn- techniques. Indeed, as underlined in Sec. \ref{sec:comparison}, computing the noise bias in these two techniques may be problematic; specifically for the case of inhomogeneous noise which has not been addressed in neither \cite{zhao_baskaran_2010} nor in \cite{kim_naselsky_2010}. Forthcoming data sets as provided by balloon-borne or ground-based experiment are plagued by {\it inhomogeneous} noise and it is therefore of primary importance to have a formul\ae~ of the noise bias applicable to inhomogeneous noise. For that purpose, we will suppose that noise in the $Q$ and $U$ maps is potentially inhomogeneous but still uncorrelated from pixel to pixel, translating into the following two-points correlation functions:
\begin{eqnarray}
	\left<n_Q(\vec{n})~n_Q(\vec{n}')\right>&=&\sigma^2_Q(\vec{n})\delta^2(\vec{n}-\vec{n}'), \nonumber \\
	\left<n_U(\vec{n})~n_U(\vec{n}')\right>&=&\sigma^2_U(\vec{n})\delta^2(\vec{n}-\vec{n}') \nonumber
\end{eqnarray}
and
\begin{equation}
	\left<n_Q(\vec{n})~n_U(\vec{n}')\right>=0. \nonumber
\end{equation}
leading to the following correlations in the harmonic space
\begin{eqnarray}
\left<n^B_{\ell m}n^{B~\star}_{\ell'm'}\right>&=&\ds\frac{1}{4}\int d\vec{n}M^2(\vec{n})\left[\sigma^2_Q(\vec{n})+\sigma^2_U(\vec{n})\right] \nonumber \\
&\times&\left[{}_{2}Y_{\ell m}(\vec{n}){}_{2}Y^\star_{\ell' m'}(\vec{n})+{}_{-2}Y_{\ell m}(\vec{n}){}_{-2}Y^\star_{\ell' m'}(\vec{n})\right]. \nonumber
\end{eqnarray}

The main challenge to compute the noise bias in the inhomogeneous case, is to find an expression of $\tilde{N}^B_\ell$ which is function of $\left<n_Q(\vec{n})~n_Q(\vec{n}')\right>$ and $\left<n_U(\vec{n})~n_U(\vec{n}')\right>$ {\it only} --as we cannot know {\it a priori} how noise is modified by taking derivatives--, and which is numerically tractable.

From the above correlation, it is easily check that on the full sky, the noise is described by a power spectrum if it is homogeneous, {\it i.e.} $\sigma^2_Q(\vec{n})=cste$ and $\sigma^2_U(\vec{n})=cste$, with
\begin{equation}
\left<n^B_{\ell m}n^{B~\star}_{\ell'm'}\right>=\frac{1}{2}\left[\sigma^2_Q+\sigma^2_U\right]\delta_{\ell,\ell'}\delta_{m,m'}. \nonumber
\end{equation}

\subsection{The \sz-technique case}
For such an approach, the noise bias is easily computed, assuming only that pixel-to-pixel correlation is vanishing. The pseudo-$a_{\ell m}$'s are the one of $W\times\chi^B$ resulting in the following noise bias of the pseudo-power spectrum :
\begin{eqnarray}
	\tilde{N}^B_\ell&=&\frac{1}{8\pi}\ds\int d\vec{n}\left(\sigma^2_Q+\sigma^2_U\right) \\
	&\times&\left(N^2_{\ell,2}W^2+4N^2_{\ell,1}\left|\partial W\right|^2+\left|\partial\partial W\right|^2\right). \nonumber
\end{eqnarray}
We refer the reader to \cite{smith_zaldarriaga_2007} or to the appendix of \cite{grain_etal_2012} for a detailed derivation of such a noise bias.

\subsection{The \zb-technique case}
The noise bias as computed in \cite{zhao_baskaran_2010} -given by their equation (50)- clearly assumes that the noise is described by a power spectrum and is therefore homogeneous. Their computation proceeds as follows. First, assume that second order moment of the noise statistics is completely described by a power spectrum, denoted $N_\ell$ in \cite{zhao_baskaran_2010}, valid on the entire celestial sphere. As a consequence, this noise bias is valid at the level of power spectra and not at the level of pseudo-power spectra. Second, compute the noise bias at the level of pseudo-$C_\ell$, denoted $\mathcal{N}_\ell$ in \cite{zhao_baskaran_2010} and denoted $\tilde{N}_\ell$ in this paper, using the convolution kernel, {\it i.e}
$$
\tilde{N}_\ell=\ds\sum_{\ell'}K_{\ell \ell'}N_{\ell'}.
$$
For the above relation to be valid, assuming that $\left<n_{\ell m}n^\star_{\ell'm'}\right>=N_{\ell}~\delta_{\ell,\ell'}~\delta_{m,m'}$ is mandatory. In other words, the noise properties should be such as the instrumental noise, as reprojected on the celestial sphere, is statistically isotropic. To our knowledge, there is no experimental set-up leading to such properties of the noise.

For the case of inhomogeneous noise, one can easily obtain the noise bias by noticing that the resulting map is equivalent to the map of the pure pseudo-$a_{\ell m}$'s as computed in the \sz-approach by replacing $W$ by $W^2$. Our purpose is to derive the noise of the power spectrum estimated from $W^2\times\chi^B$ as a function of the noise power per pixel of the $Q$ and $U$ maps. Our starting point is the pseudo-multipoles given by :
\begin{equation}
	\tilde{\chi}^B_{\ell m}=\ds\int W^2\times\chi^B\times Y^\star_{\ell m} d\vec{n}.
\end{equation}
Since the $\chi^B$ field is defined by $\chi^B=i\left[\bar\partial\bar\partial P_2-\partial\partial P_{-2}\right]/2$, it is easily shown by performing two integrations by parts and using the fact that $W^2$ and $\partial(W^2)$ are vanishing at the contour, that :
\begin{equation}
	\tilde{\chi}^B_{\ell m}=\frac{i}{2}\ds\int \left[P_2\bar\partial\bar\partial(W^2Y^\star_{\ell m})-P_{-2}\partial\partial(W^2Y^\star_{\ell m})\right] d\vec{n},
	\label{equ:appchi}
\end{equation}
which exactly is the definition of the pure pseudo-multipoles. As a consequence, the noise bias for the \zb-method is given by the noise bias as computed in the \sz-method. 

However, $W^2\chi^B$ is effectively computed using the following expression :
\begin{eqnarray}
	W^2(\vec{n})\chi^B(\vec{n})&=&\left(\frac{i}{2}\right)W\left[\bar\partial\bar\partial\left(WP_{2}\right)-\partial\partial\left(WP_{-2}\right)\right]  \\
	&-&i\left[\bar\partial W\times\bar\partial\left(WP_{2}\right)-\partial W\times\partial\left(WP_{-2}\right)\right] \nonumber \\
	&-&\left(\frac{i}{2}\right)W\left[\left(\bar\partial\bar\partial W\right)P_{2}-\left(\partial\partial W\right)P_{-2}\right]. \nonumber \\
	&+&i\left[{\left(\bar\partial W\right)^2}\times P_{2}-{\left(\partial W\right)^2}\times P_{-2}\right]. \nonumber
\end{eqnarray}
Let us show positively that we obtain the same expression for $\tilde{\chi}^B_{\ell m}$ using the right-hand-side of the above formul\ae. By plugging the above expression into the expression of $\tilde{\chi}^B_{\ell m}$, one can perform some integration by parts in order to replace terms like $F\times\partial(WP_{\pm2})$ by terms like $(WP_{\pm2})\times\partial F$. For the first line of the rhs of the above expression, two integrations by parts are required and only one is needed for the second line. With such a procedure, some contour integrals should appear. However, the integrant in those contour integrals are always proportionnal to $W$ and/or $\partial W$. As those functions are required to vanish at the contour, all the contour integrals are equal to zero and we are left with 
\begin{equation}
	\tilde{\chi}^B_{\ell m}=\frac{i}{2}\ds\int \left[P_2\times{}_{2}\mathcal{W}^\star_{\ell m}-P_{-2}\times{}_{-2}\mathcal{W}^\star_{\ell m}\right]d\vec{n},
\end{equation}
with
\begin{eqnarray}
	{}_{2}\mathcal{W}^\star_{\ell m}&=&W\times\bar\partial\bar\partial(WY^\star_{\ell m})+2 W\times\bar\partial(Y^\star_{\ell m}\bar\partial W) \\
	&-&W\times Y^\star_{\ell m}\times\bar\partial\bar\partial W+2\times Y^\star_{\ell m}\times\left(\bar\partial W\right)^2, \nonumber \\
	{}_{-2}\mathcal{W}^\star_{\ell m}&=&W\times\partial\partial(WY^\star_{\ell m})+2 W\times\partial(Y^\star_{\ell m}\partial W) \\
	&-&W\times Y^\star_{\ell m}\times\partial\partial W+2\times Y^\star_{\ell m}\times\left(\partial W\right)^2. \nonumber
\end{eqnarray}
Expanding the derivatives and re-arranging appropriately the different terms, one easily sees that $\tilde{\chi}^B_{\ell m}$ is given by \eref{equ:appchi}. From Eq. \eref{equ:appchi} and defining the noise bias as
\begin{equation}
	\tilde{N}^B_\ell=\frac{1}{2\ell+1}\ds\sum_{m=-\ell}^\ell \left<\left|\tilde{\chi}^B_{\ell m}\right|^2\right> \nonumber
\end{equation}
with $\tilde{\chi}^B_{\ell m}$ containing noise only, it is straightforward to apply the noise bias calculation performed for the pure pseudo-$C_\ell$ techniques (see {\it e.g.} appendix C of \cite{grain_etal_2012}) to get:
\begin{eqnarray}
	\tilde{N}^B_\ell&=&\frac{1}{8\pi}\ds\int d\vec{n}\left(\sigma^2_Q+\sigma^2_U\right) \\
	&\times&\left(N^2_{\ell,2}W^4+4N^2_{\ell,1}\left|\partial W^2\right|^2+\left|\partial\partial W^2\right|^2\right). \nonumber
\end{eqnarray}

\subsection{The \kn-method case}
For such a method, computing the noise bias is more involved. We list here three possible approaches but none of them allows a numerical calculation on $\tilde{N}^B_\ell$ from $\sigma_{Q/U}$ to be implemented. 

(i)~{\it First method}: The pseudo-$a_{\ell m}$'s are derived via
\begin{equation}
	\tilde{\chi}^B_{\ell m}=\ds\int M_{\chi^B}\times\tilde\chi^B\times Y^\star_{\ell m}d\vec{n},
	\label{eq:chielln}
\end{equation}
with
\begin{equation}
	\tilde\chi^B=\frac{i}{2}\left[\bar\partial\bar\partial(W\times P_2)-\partial\partial(W\times P_{-2}\right].
	\label{eq:chin}
\end{equation}
The noise bias can therefore be expressed as a function of the correlation function of $\tilde\chi^B$, denoted $\mathcal{C}_{\chi}(\vec{n},\vec{n}')\equiv\left<\tilde\chi^B(\vec{n})\tilde\chi^B(\vec{n}')\right>$, inserting only noise in $P_{\pm2}$ for computing such a correlation function:
\begin{eqnarray}
	\tilde{N}^B_{\ell}&=&\ds\iint_{4\pi}M_{\chi^B}(\vec{n})\times M_{\chi^B}(\vec{n}')\times\left<\tilde\chi^B(\vec{n})\tilde\chi^B(\vec{n}')\right> \nonumber \\
	&\times&\sum_{m=-\ell}^\ell\frac{Y^\star_{\ell m}(\vec{n})Y_{\ell m}(\vec{n}')}{2\ell+1}~d\vec{n}~d\vec{n}'.
\end{eqnarray}
By using the definition of $\tilde\chi^B$ as a convolution, {\it i.e.}
\begin{eqnarray}
	\tilde\chi^B(\vec{n})&=&\ds\int W(\vec{n}')\left[F_+(\vec{n},\vec{n}')P_{2}(\vec{n}')\right. \nonumber \\
	&-&\left.F_-(\vec{n},\vec{n}')P_{-2}(\vec{n}')\right]d\vec{n}', \nonumber
\end{eqnarray}
and the fact that for noise
\begin{eqnarray}
	\left<P_{\pm2}(\vec{n})P_{\mp2}(\vec{n}')\right>&=&\left(\sigma^2_Q(\vec{n})+\sigma^2_U(\vec{n})\right)\delta(\vec{n}-\vec{n}') \nonumber \\
	\left<P_{\pm2}(\vec{n})P_{\pm2}(\vec{n}')\right>&=&\left(\sigma^2_Q(\vec{n})-\sigma^2_U(\vec{n})\right)\delta(\vec{n}-\vec{n}') \nonumber
\end{eqnarray}
it is shown that\footnote{We remind that we fix the noise in the $Q$ map not to be correlated to the noise in the $U$ maps, and that $F_-$ is the complex conjugate of $(-F_+)$.} 
\begin{eqnarray}
\mathcal{C}_{\chi}(\vec{n},\vec{n}')&=&\ds\int W(\vec{n}'')\left\{\left[\sigma^2_Q(\vec{n}'')+\sigma^2_U(\vec{n}'')\right]\mathcal{F}^+_{\vec{n},\vec{n}'}(\vec{n}'')\right. \nonumber \\
	&+&\left.\left[\sigma^2_Q(\vec{n}'')-\sigma^2_U(\vec{n}'')\right]\mathcal{F}^-_{\vec{n},\vec{n}'}(\vec{n}'')\right\}d\vec{n}'' \nonumber
\end{eqnarray}
with
\begin{eqnarray}
	\mathcal{F}^+_{\vec{n},\vec{n}'}(\vec{n}'')&=&2\mathrm{Re}\left[F_+(\vec{n},\vec{n}'')F^\star_+(\vec{n}',\vec{n}'')\right], \\
	\mathcal{F}^-_{\vec{n},\vec{n}'}(\vec{n}'')&=&2\mathrm{Re}\left[F_+(\vec{n},\vec{n}'')F_+(\vec{n}',\vec{n}'')\right].
\end{eqnarray}
The above correlation function cannot be further simplified unless assuming full-sky coverage and the noise to be homogeneous. As a consequence, computing the noise bias directly from the noise properties of the Stokes parameters maps is numerically prohibitive (at least in the not-so-general case of inhomogeneous noise).

(ii)~{\it Second method}: A possible way out --inspired by the computation of the noise bias in the \zb-approach-- is to start from Eqs. \eref{eq:chielln} \& \eref{eq:chin} and subsequently perform two integrations by parts in order to {\it transfer} the derivative operators from $M\times P_{\pm2}$ to $M_{\chi^B}\times Y^\star_{\ell m}$. On defining $\Omega_{\chi^B}$ and $\partial\Omega_{\chi^B}$ the portion of the sky and the contour of such a portion defined by the binary mask $M_{\chi^B}$, we are lead to evaluate one integral on $\Omega_{\chi^B}$ (denoted the {\it domain-integral}) and two other integrals on $\partial\Omega_{\chi^B}$ (denoted {\it contour-integrals}). The integrant of the the domain-integral is 
\begin{equation}
\left(\frac{i}{2}\right)W\left[P_{2}\times\bar\partial\bar\partial(M_{\chi^B}Y^\star_{\ell m})-P_{-2}\times\partial\partial(M_{\chi^B}Y^\star_{\ell m})\right]. \nonumber
\end{equation}
Because $M_{\chi^B}$ is constant-valued on the domain and because $M_{\chi^B}\times W=M_{\chi^B}$ --the domain covered by $M_{\chi^B}$ is at most the sub-part of the domain covered by $W$ such as $W=cste=1$--, the integrand is simply given by
\begin{equation}
\left(\frac{i}{2}\right)M_{\chi^B}\left[P_{2}\times\bar\partial\bar\partial(Y^\star_{\ell m})-P_{-2}\times\partial\partial(Y^\star_{\ell m})\right], \nonumber
\end{equation}
However, the two integrands relative to the contour-integrals are of the form
\begin{equation}
\left(\frac{i}{2}\right)\left[\bar\partial(W P_2)-\partial(W P_{-2})\right]M_{\chi^B} Y^\star_{\ell m}, \nonumber
\end{equation}
and
\begin{equation}
\left(\frac{i}{2}\right)W\left[P_2\times\bar\partial(M_{\chi^B} Y^\star_{\ell m})- P_{-2}\times\partial(M_{\chi^B} Y^\star_{\ell m})\right]. \nonumber
\end{equation}
Because $M_{\chi^B}=1$ and $W=1$ on $\partial\Omega_{\chi^B}$, the two contour-integrals are {\it not} vanishing. Though the second contour-integral can be expressed as functions of $\left<n_Q(\vec{n})~n_Q(\vec{n}')\right>$ and $\left<n_U(\vec{n})~n_U(\vec{n}')\right>$, the first contour-integral is still a function of the derivative of $P_{\pm2}$ preventing us from computing the noise bias.

(iii)~{\it Third method}: This issu of contour-integrals can be naturally circumvented by replacing the binary mask $M_{\chi^B}$ by an apodized mask, $W_{\chi^B}$, satisfying the Dirichlet and Neumann boundary conditions. With such a trick and using that $W=1$ on $\Omega_{\chi^B}$, the multipoles of $\tilde\chi^B$ are subsequently given by 
\begin{equation}
	\tilde{\chi}^B_{\ell m}=\frac{i}{2}\ds\int \left[P_{2}\times\bar\partial\bar\partial(W_{\chi^B}Y^\star_{\ell m})-P_{-2}\times\partial\partial(W_{\chi^B}Y^\star_{\ell m})\right]. \nonumber
\end{equation}
However, the above-defined pseudo-multipoles are no more than the definition of the {\it pure} pseudo-multipoles, but now computed on a reduced domain. In other word, replacing $M_{\chi^B}$ by an appropriately apodized window function makes the \kn-method to reduce to the \sz- and \zb-techniques but on a {\it smaller} portion of the sky. With such an implementation of the \kn-approach, part of the information is inherently lost as compared to the two other approaches and there would be no reason to use the \kn-method.



\begin{thebibliography}{99}

\bibitem{zaldarriaga_seljak_1997}
	M. Zaldarriaga \&\  U. Seljak, \prd~ {\bf55} 1830 (1997)
\bibitem{kamionkowski_etal_1997}
	M. Kamionkowski, A. Kosowsky \& A. Stebbins, Phys. Rev. Lett. {\bf 78} 2058 (1997)
\bibitem{dasi}
	E. M. Leitch {\it et al.}, \nat~{\bf420} 763 (2002); J. M. Kovac {\it et al.}, \nat~{\bf420} 772 (2002)
\bibitem{wmap_pol}
	C. L. Bennett {\it et al.}, \apj~Supp. {\bf148} 1 (2003); G. Hinshaw {\it et al.}, \apj~Supp. {\bf170} 228 (2007); E. Komatsu {\it et al.}, \apj~Supp. {\bf192} 18 (2011)
\bibitem{quad_pol}
	M. L. Brown {\it et al.}, \apj~{\bf705} 978 (2009)
\bibitem{bicep_pol}
	H. C. Chiang {\it et al.}, \apj~{\bf711} 1123 (2010)
\bibitem{planck}
	{\sc planck}: \\
	{\tt http://www.esa.int/esaSC/120398\_index\_0\_m.html}
\bibitem{seljak_zaldarriaga_1997}
	U. Seljak \&\ M. Zaldarriaga, \prl~{\bf78} 2054 (1997)
\bibitem{spergel_zaldarriaga_1997}
	D. N. Spergel \&\ M. Zaldarriaga, \prl~{\bf79} 2180 (1997)
\bibitem{zaldarriaga_1997}
	M. Zaldarriaga, \prd~{\bf55 1822}
\bibitem{zaldarriaga_seljak_1998}
	M. Zaldarriaga \&\ U. Seljak, \prd {\bf 58} 023003 (1998)
\bibitem{polarbear}
	{\sc polarbear}: \\
	{\tt http://bolo.berkeley.edu/polarbear/?q=science}
\bibitem{sptpol}
	{\sc sptpol}: \\
	{\tt http://pole.uchicago.edu/}
\bibitem{qubic}
	{\sc qubic}: \\
	{\tt http://www.qubic-experiment.org/}
\bibitem{actpol}
	{\sc actpol}: \\
	{\tt http://www.princeton.edu/act/}
\bibitem{ebex}
	{\sc ebex}: \\
	{\tt http://groups.physics.umn.edu/cosmology/ebex/}
\bibitem{spider}
	{\sc spider}: \\
	{\tt http://cmb.phys.cwru.edu/ruhl\_lab/spider.html}
\bibitem{litebird}
	{\sc l}ite\textsc{bird}: \\
	{\tt http://cmbpol.kek.jp/litebird/}
\bibitem{bpol_website}
	{\sc co}r\textsc{e}: \\ 
	{\tt http://www.core-mission.org/}
\bibitem{pixie}
	\textsc{pix}i\textsc{e}: \\
	A. Kogut {\it et al.}, JCAP {\bf 07} 025 (2011)
\bibitem{zaldarriaga_2001}
	M. Zaldarriaga, \prd~{\bf64} 103001 (2001)
\bibitem{hauser_peebles_1973}
	M.G. Hauser, \&\ P.J.E. Peebles, \apj {\bf185} {757} (1973)
\bibitem{hansen_gorski_2003}
	F. Hansen \&\ K. M. G\'orski, MNRAS {\bf 343} 559 (2003)
\bibitem{hivon_etal_2002}
	E. Hivon, K. M. G\'orski, C. B. Netterfield, B. P. Crill, S. Prunet, \&\ F. Hansen, \apj {\bf567} {2} (2002)
\bibitem[Smith (2006)]{smith_2006}
	K. M. Smith, \prd {\bf74} 083002 (2006)
\bibitem[Smith \etal~(2007)]{smith_zaldarriaga_2007}
	K. M. Smith \&\ M. Zaldarriaga, \prd~ {\bf76} 0043001 (2007)
\bibitem[Grain \etal~(2009)]{grain_etal_2009}
	J. Grain, M. Tristram \&\ R. Stompor, \prd~ {\bf79} 123515 (2009)
\bibitem[Zhao \etal~(2010)]{zhao_baskaran_2010}
	W. Zhao \&\ D. Baskaran, \prd~ {\bf82} 023001 (2010)
\bibitem[Kim \etal~(2010)]{kim_naselsky_2010} 
	J. Kim \&\ P. Naselsky, \aap~ {\bf519} A104 (2010)
\bibitem[Kim (2011)]{kim_2010}
	J. Kim, \aap~ {\bf531} A32 (2011)
\bibitem[Bowyer \etal~(2011)]{bowyer_2011}
	J. Bowyer, A. Jaffe \&\ D. I. Novikov, arXiv:1101.0520 [astro-ph.CO]
\bibitem{tristram_etal_2005}
	M. Tristram, J. F. Mac\'\i{a}s-P\'erez, C. Renault \&\ D. Santos, MNRAS {\bf358} 833 (2005)
\bibitem[Bock \etal~(2008)]{bock_etal_2008}
	J. Bock {\it et al.}, arXiv:0805.4207 [astro-ph]
\bibitem{britt_etal_2010}
	B. Reichborn-Kjennerud, {\it et al.}, Proc. SPIE Int. Soc. Opt. Eng. {\bf7741} {77411C} (2010) 
\bibitem[Bunn \etal~(2003)]{bunn_etal_2003}
	E. F. Bunn, M. Zaldarriaga, M. Tegmark \&\ A. de Oliveira-Costa, \prd~ {\bf67} 023501 (2003)
\bibitem[Grain \etal~(2012)]{grain_etal_2012}
	J. Grain, M. Tristram \&\ R. Stompor, \prd~ {\bf86} 076005 (2012)
\bibitem{hinshaw_etal_2003}
	G. Hinshaw {\it et al.}, Astrophys. J. Supp. {\bf148} {135} (2003)
\bibitem{wmap_7yr}
	{\sc wmap}: \\
	{\tt http://lambda.gsfc.nasa.gov/product/map/current}
\bibitem{larson_etal_2011}
	D. Larson {\it et al.}, Astrophys. J. Supp. {\bf192} 16 (2011)
\bibitem{s2hat}{\sc s$^2$hat}:\\
	{\tt http://www.apc.univ-paris7.fr/APC\_CS/Recherche/\\Adamis/MIDAS09/software/s2hat/s2hat.html}
\bibitem{pures2hat}{\sc pureS$^2$HAT}:\\
	{\tt http://www.apc.univ-paris7.fr/APC\_CS/Recherche/\\
	Adamis/MIDAS09/software/pures2hat/pureS2HAT.html} 
\bibitem{hupca_etal_2010}
	I. O. Hupca, J. Falcou, L. Grigori \& R. Stompor, Lecture Notes in Computer Science {\bf 7155} 355 (2012)
\bibitem{szydlarski_etal_2011}
         M. Szydlarski, P. Esterie, J. Falcou, L. Grigori, \& R. Stompor, arXiv:11060159 [c.:DC]
\bibitem{gorski_etal_2005}
	K. M. G\'orski, E. Hivon, A. J. Banday, B. D. Wandelt, F. K. Hansen, M. Reinecke \& M. Bartelmann, \apj {\bf622} {759} (2005) 
\bibitem{lewis_etal_2000}
	A. Lewis, A. Challinor \& A. Lasenby, \apj {\bf538} {473} (2000)

\end{thebibliography}
\end{document}